\documentclass[a4paper]{article}

\usepackage{hyperref}
\usepackage[a4paper, top=25mm, left=10mm, right=10mm, bottom=30mm, headsep=10mm, footskip=12mm]{geometry}
\usepackage{authblk}

\usepackage[nointlimits]{amsmath}
\usepackage{siunitx}

\usepackage{caption}
\usepackage{graphicx}
\usepackage{subfigure}
\usepackage{xcolor}
\usepackage{lscape}
\usepackage{mathtools}

\usepackage[ruled,vlined]{algorithm2e}
\usepackage{algorithmic}

\definecolor{mycolor_blue}{RGB}{66,124,161}
\definecolor{mycolor_grey}{RGB}{198,198,198}

\usepackage{tikz}
\usepackage{pgfplots}

\usetikzlibrary{patterns}

\usetikzlibrary{decorations.pathmorphing,arrows,intersections}

\pgfplotsset{
	kurze Legende/.style={
		legend image code/.code={
			\draw[##1,mark repeat=2,line width=0.6pt]
			plot coordinates {
				(0cm,0cm)
				(0.3cm,0cm)
			};
		}
	}
}

\pgfplotsset{
	compat = newest,
	scale only axis, 
	max space between ticks = 50pt,
	ticklabel style = {font=\footnotesize},
	legend style =  {font=\footnotesize},
	grid = major,
	grid style = {dotted},
	legend columns=1, 
	xtick pos=left,
	ytick pos=left
}

\newcommand{\analytiSolutionPictures}{
	\pgfplotsset{  
		width=0.24\textwidth,
		height=0.25\textwidth,
		ylabel style={text width=0.2\textwidth,align=center}
	}
}

\pgfplotsset{select coords between index/.style 2 args={
		x filter/.code={
			\ifnum\coordindex<#1\fi
			\ifnum\coordindex>#2\fi
		}
}}

\definecolor{color1}{HTML}{0060AD} 
\definecolor{color2}{HTML}{FF4500} 
\definecolor{color3}{HTML}{FFA500} 
\definecolor{color4}{HTML}{006400} 
\definecolor{color5}{HTML}{9400D3} 
\definecolor{color6}{HTML}{800000} 
\definecolor{color7}{HTML}{000000} 
\definecolor{color8}{HTML}{0000FF} 
\definecolor{color9}{HTML}{FF0000} 
\definecolor{mycolor_blue}{RGB}{66,124,161}
\definecolor{mycolor_grey}{RGB}{198,198,198} 

\tikzstyle{line1} = [color=color7,semithick] 
\tikzstyle{line2} = [color=color2,densely dotted,semithick]
\tikzstyle{line3} = [color=color1,densely dashed,semithick]
\tikzstyle{line4} = [color=color5,dash dot,semithick]
\tikzstyle{line5} = [color=color4,dash dot dot,semithick]
\tikzstyle{line6} = [color=color6,semithick]

\tikzstyle{mark1} = [color=color7,mark=x,mark size=2pt,mark options=solid,semithick] 
\tikzstyle{mark2} = [color=color2,mark=o,mark size=2pt,mark options=solid,semithick]
\tikzstyle{mark3} = [color=color1,mark=*,mark size=2pt,mark options=solid,semithick]
\tikzstyle{mark4} = [color=color5,mark=triangle,mark size=2pt,mark options=solid,semithick]
\tikzstyle{mark5} = [color=color4,mark=square,mark size=2pt,mark options=solid,semithick]

\title{
Adjoint Complement to the Universal Momentum Law of the Wall 
}

\author{Niklas K\"uhl\thanks{niklas.kuehl@tuhh.de}, Peter M. M\"uller and Thomas Rung}

\affil{Hamburg University of Technology, Institute for Fluid Dynamics and Ship Theory, Am Schwarzenberg-Campus 4, D-21075 Hamburg, Germany}

\date{\today}

\begin{document}

\maketitle

\begin{abstract}
The paper is devoted to an adjoint complement to the universal Law of the Wall (LoW) for fluid dynamic momentum boundary layers. The latter typically follows from a strongly simplified, unidirectional shear flow under a constant stress assumption.
We first derive the adjoint companion of the simplified momentum equation, while distinguishing between two strategies. Using mixing-length arguments, we demonstrate that the frozen turbulence strategy and a LoW-consistent (differentiated) approach provide virtually the same adjoint momentum equations, that differ only in a single scalar coefficient controlling the inclination in the logarithmic region.
Moreover, it is seen that an adjoint LoW can be derived which resembles its primal counterpart in many aspects.
%
The  strategy is also compatible with wall-function assumptions for prominent RANS-type two-equation turbulence models, which ground on the mixing-length hypothesis.

As a direct consequence of the frequently employed assumption that all primal flow properties algebraically scale with the friction velocity, it is demonstrated that a simple algebraic expression provides a consistent closure of the adjoint momentum equation in the logarithmic layer.
%
%
This algebraic adjoint closure might also serve as an approximation for  more general adjoint flow optimization studies using standard one- or two-equation Boussinesq-viscosity models for the primal flow. 
%
Results obtained from the suggested algebraic closure are verified against the primal/adjoint LoW formulations for both, low- and high-Re settings.
Applications included in this paper refer to two- and three-dimensional shape optimizations of internal and external engineering flows. Related results indicate that 
the proposed adjoint algebraic  turbulence closure accelerates the optimization process and provides improved optima at no computational surplus in comparison to the frozen turbulence approach.
\end{abstract}


\section{Introduction} 
\label{sec:Introduction}
This paper is concerned with the formulation of an adjoint law of the wall (LoW) serving the formulation of momentum boundary conditions in an adjoint analysis and a related algebraic 
treatment of turbulence in the adjoint framework.
In the context of local fluid dynamic optimization, the adjoint analysis aims at the efficient computation of derivative information for an integral objective functional with respect to (w.r.t) a general control function \cite{giles1997adjoint, giles2000introduction, papoutsis2016continuous, kroger2018adjoint, kapellos2019unsteady}.
In continuous space, the dual or adjoint flow state can be interpreted as a co-state and always follows from the underlying primal Partial Differential Equation (PDE) governed model that describes the flow physics. 
However, the appropriate formulation of boundary conditions is often not intuitively clear in a PDE-based, continuous adjoint framework and the development of numerical strategies clearly lags behind the primal progress \cite{soto2004computation, othmer2008continuous, zymaris2010adjoint, stuck2013adjoint, othmer2014adjoint}.  

Modelling equations for the turbulent closure appear comparatively complex already on the primal side. The latter is underlined by an unfavourable algorithmic complexity that contains possibly non-differentiable expressions making it unhandy for a continuous adjoint approach which has motivated the neglect of adjoint turbulence models in line with the frozen turbulence approach \cite{soto2004adjoint, othmer2008continuous, dwight2006effects}.
However, the influence of the variation of the turbulence parameters is an open discussion \cite{marta2013handling, dwight2006effects} which is why discrete adjoint approaches using automatic differentiation have been derived that aim at a synchronization of the primal and dual turbulent development states, cf. \cite{nielsen2004implicit, nielsen2010discrete, nielsen2013discrete}.
The discrete approach passes over the adjoint PDE and directly bridges the discrete linearized primal flow into a consistent discrete dual approach, cf. \cite{giles1997adjoint, giles2000introduction} or \cite{vassberg2006aerodynamic_II, vassberg2006aerodynamic}. Despite the various merits and drawbacks of the discrete vs. the continuous adjoint method, the latter is unique for its invaluable contribution to a physical understanding.
The development of the continuous adjoint method with respect to adjoint turbulence modelling initially started with the derivation of adjoint one equation closures \cite{zymaris2009continuous, bueno2012continuous, bagheri2020adjoint} followed by the complete linearization of prominent statistical closures, e.g. an adjoint $k-\varepsilon$ \cite{papoutsis2015continuous, zymaris2010adjoint} and $k-\omega$ \cite{kavvadias2015continuous, hartmann2011adjoint, manservisi2016numerical, manservisi2016optimal} model.
All previously mentioned contributions share the idea of deriving adjoint turbulence modelling equations. 
Optimizations of complex engineering flows using fully consistent, differentiated turbulence transport models are however rare. Primal turbulence transport models inhere multiple non-linearities and inter-parameter couplings, that significantly hamper the robustness and the efficiency of a consistent adjoint framework  and hinder their utilization in engineering applications. On the other hand, the continuous adjoint framework gives access to dedicated adjoint turbulence modelling at a lower level of adjoint consistency. Thus, one research question of the present effort is to  investigate the potential of an algebraic adjoint turbulence treatment that offers the algorithmic benefits of a frozen turbulence approach.

In contrast to former studies, our study originates from analysing the adjoint complement to a simple unidirectional turbulent shear flow, which is the foundation of virtually all wall function based turbulent boundary conditions using the mixing-length hypothesis \cite{prandtl25, pope2001turbulent}. 
%
We distinguish between two adjoint turbulence formulations, i.e. an algebraic, mixing-length based approach and a simple frozen turbulence approach.
With reference to the adjoint LoW, both formulations differ only in a single scalar coefficient in the logarithmic region
and a simple scaling with the ratio of the friction velocities.
The analysis suggests a surprisingly simple algebraic  approximation for an adjoint turbulence treatment. Results obtained by this strategy are deemed consistent to LoW physics and indicate 
improvements over the frozen turbulence assumption when applied to more general flows without solving an adjoint turbulence transport model.

The remainder of the paper is organised as follows: Section \ref{sec:derivation} and \ref{sec:adderivation} are concerned with the derivation of the adjoint unidirectional shear flow equations for a frozen as well as a consistently linearized turbulent viscosity contribution. The 
subsequent section \ref{sec:adjoint_low} derives an adjoint complement to the primal LoW. In section \ref{sec:adjoint_wall_functions} we discuss our findings w.r.t. more sophisticated two-equation turbulence models. Verification studies are presented in the \ref{sec:numerical_results}th section. 
Section \ref{sec:appl} scrutinizes the performance of the suggested algebraic model for several internal and external shape optimization examples of engineering relevance. The final section \ref{sec:conclusion} provides conclusions and outlines future research. Within the publication, Einstein’s summation convention is used for lower-case Latin subscripts and vectors as well as tensors are defined with reference to Cartesian coordinates.

%

\section{Primal Unidirectional Shear Flow}
\label{sec:derivation}
We start with a brief discussion of a simple  --yet commonly used-- incompressible primal flow description. 
The discussion is confined to plane wall flows, using a local orthogonal coordinate system as illustrated in Fig. \ref{fig:channel_flow}, where $y$ denotes the wall normal coordinate or distance and $x$ refers to the wall tangential direction. 
The flow field is usually considered to be fully developed and assumed as uni-directional, i.e. $u(y)$ in the vicinity of the wall.
Extensions to more general curved near wall flows have been published in \cite{zymaris2010adjoint, rung2001universal} but are not considered here to save space.
A key element of the concept - which is crucial for the formulation of boundary conditions for turbulent wall flows -- is the constant shear stress hypothesis. The latter assumes $\tau_\mathrm{eff} = \mathrm{const.}$ for the inner region of a wall boundary layer $y/\Delta <<1$ where $y=\Delta$ denotes the outer edge of the boundary layer. The simple relation substitutes the momentum equation above the wall and supports the derivation of both the primal and the adjoint LoW, viz.
%
%
%
%
\begin{align}
\mathrm{R}^\mathrm{u}: \qquad \frac{\mathrm{d}\tau_\mathrm{eff}}{\mathrm{d} y}  =   \frac{\mathrm{d}}{\mathrm{d} y} \left[ \mu_\mathrm{eff} \frac{\mathrm{d} u}{\mathrm{d} y} \right] = 0  \, , \qquad \mathrm{with} \qquad \mu_\mathrm{eff} = \mu + \mu_\mathrm{t} \; .
\label{equ:primal_channel_equation}
\end{align}
The validity of (\ref{equ:primal_channel_equation}) is restricted to approximately the inner 20\% of the boundary layer and widens with increasing boundary-layer thickness, cf. \cite{pope2001turbulent, wilcox1998turbulence}. 
An isotropic Boussinesq-viscosity model (BVM) is frequently employed in the majority of Reynolds-averaged Navier-Stokes (RANS) or large-eddy simulation (LES) frameworks to supplement the laminar, molecular stress $\tau_\mathrm{l}= \mu \;  \mathrm{d} u/\mathrm{d} y$ by a companion turbulent stress  $\tau_\mathrm{t}= \mu_\mathrm{t} \;  \mathrm{d} u/\mathrm{d} y$ and  close the formulation. Mind that despite the particular turbulence model employed to determine $\mu_\mathrm{t}$, e.g. the $k-\epsilon$, $k-\omega$ or $\nu_\mathrm{t}$ formulation 
\cite{wilcox1998turbulence,spalart1992one}, their values usually comply with the mixing length hypothesis in the logarithmic layer, i.e. $\mu_\mathrm{t} = \rho \left( \kappa \, y \right)^2 \mathrm{d} u/\mathrm{d} y$, where ($\kappa y$)  denotes the mixing length and $\kappa$ is the von-Karman constant.


%

\section{Adjoint Unidirectional Shear Flow}
\label{sec:adderivation}
The adjoint system studied herein should provide gradient information for a boundary-based objective $ j_\mathrm{\Gamma} $ w.r.t a general control parameter, e.g. the shape of the wall ($\delta_\mathrm{y} j_\mathrm{\Gamma}$).
%
%
A widely used exemplary 
objective refers to the flow induced shear force $j_{\mathrm{\Gamma}} =  \mu_{\mathrm{eff}} [ \mathrm{d} u / \mathrm{d} y]$ along the wall. We would like to point out that there are different adjoint answers to the same question, e.g. regarding fluid flow-induced forces \cite{kuhl2019decoupling, kuhl2020continuous}.
If attention is given to a boundary layer, e.g. the lower half of a channel outlined in Fig.
\ref{fig:channel_flow}, the constraint optimization problem 
 is transformed into an unconstrained formulation based on a Lagrangian $L$ 
\begin{align}
    \mathrm{min} \quad 
    J =  \mu_\mathrm{eff} \frac{\mathrm{d} u}{\mathrm{d} y} \bigg \vert_\mathrm{w}  
    \quad \mathrm{s.t.} \quad \mathrm{R}^\mathrm{u} = 0 \qquad \rightarrow \qquad 
    L = J + \int \hat{u} \, \mathrm{R}^\mathrm{u} \, \mathrm{d} y \, , 
    \label{eq:anfang}
\end{align}
where the index $(\cdot )_\mathrm{w}$ denotes to a wall value.
Equation (\ref{eq:anfang}) inheres a Lagrangian multiplier $\hat{u}$ which  is frequently labeled as the dual or adjoint velocity. Its dimension depends on the underlying objective, e.g. $[\hat{u}] = [J]/( [R^\mathrm{u}] \, m^2)$ where $[J] = [j_\mathrm{\Gamma}] \, m$ represents the units of the boundary-based objective.
The total variation of the Lagrangian leads to the adjoint equation. Using $\rho \nu = \mu$ together with a constant density $\rho$, we obtain
\begin{align}
    \delta L = 
    (\delta  \nu_\mathrm{eff}) \frac{\mathrm{d} u}{\mathrm{d} y}\bigg \vert_{w} + \nu_\mathrm{eff} \frac{\mathrm{d} (\delta u)}{\mathrm{d} y}\bigg \vert_{w} + \int \hat{u} \left[ \frac{\mathrm{d}}{\mathrm{d} y} \left[ (\delta  \nu_\mathrm{eff}) \frac{\mathrm{d} u}{\mathrm{d} y} + \nu_\mathrm{eff} \frac{\mathrm{d} (\delta u)}{\mathrm{d} y} \right] \right] \, \mathrm{d} y \, .
\end{align}

\paragraph{The frozen turbulence assumption} neglects the variation of the turbulent viscosity, i.e. $\delta \nu_\mathrm{eff} = 0$. 
An isolation of $\delta u$ allows the formulation of first order optimality conditions, viz.
\begin{align}
    \delta_\mathrm{u} L \cdot \delta u= 
    \nu_\mathrm{eff} \frac{\mathrm{d} (\delta u)}{\mathrm{d} y} \bigg \vert_{w} +  \left[\nu_\mathrm{eff} \left( \hat{u} \frac{\mathrm{d} (\delta u)}{\mathrm{d} y} - \frac{\mathrm{d} \hat{u}}{\mathrm{d} y} (\delta u) \right) \right]_{w}^{\Delta} + \int \delta u \left[ \frac{\mathrm{d}}{\mathrm{d} y} \left[ \nu_\mathrm{eff} \frac{\mathrm{d} \hat{u} }{\mathrm{d} y} \right] \right] \, \mathrm{d} y 
    \qquad \overset{!}{=} 0 \qquad \forall \, \delta u \label{equ:frozen_turb_variation} \, .
\end{align}
Here $y=\Delta$ marks the position of the outer boundary. 
The adjoint equation to (\ref{equ:primal_channel_equation}) follows from the integral expression in (\ref{equ:frozen_turb_variation}) and reads
\begin{align}
\mathrm{\hat{R}}^\mathrm{\hat{u}, F}&: \qquad  \frac{\mathrm{d}}{\mathrm{d} y} \left[  (\nu + \nu_\mathrm{t}) \frac{\mathrm{d} \hat{u}}{\mathrm{d} y} \right] = 0 \label{equ:frozen_turbadjoint_equation} \, .
\end{align}
The asterisk (F) indicates the adjoint equation based on the frozen turbulence assumption that resembles its primal counterpart in a self-adjoint manner.
The boundary conditions along the wall  
as well as the outer boundary 
follow from the remaining terms, viz.
\begin{alignat}{4}
    y=\Delta&: \qquad \left[ \hat{u} \frac{\mathrm{d} (\delta u)}{\mathrm{d} y} - \frac{\mathrm{d} \hat{u}}{\mathrm{d} y} (\delta u) \right] \qquad \mathrm{with} \qquad \delta \left( \frac{\mathrm{d} u}{\mathrm{d} y} \right) = \frac{\mathrm{d} (\delta u)}{\mathrm{d} y} = 0 \qquad &&\rightarrow \qquad \frac{\mathrm{d} \hat{u}}{\mathrm{d} y} \bigg \vert_{\Delta} = 0 \label{equ:adjoint_bc_1} \\
    y=0&: \qquad \left[ (1+\hat{u}) \frac{\mathrm{d} (\delta u)}{\mathrm{d} y}  - \frac{\mathrm{d} \hat{u}}{\mathrm{d} y} (\delta u) \right] \qquad \mathrm{with} \qquad \delta u = 0
    \qquad &&\rightarrow \qquad \hat{u} \big \vert_{\mathrm{w}} = -1 \label{equ:adjoint_bc_2}.
\end{alignat}

\paragraph{A consistent approach} 
also considers the variation of the turbulent viscosity. Thanks to the employed mixing length hypothesis, the turbulent viscosity exclusively depends on the tangential mean velocity and the related variation reads $\delta \nu_\mathrm{eff} = (\kappa y)^2 \, (\mathrm{d} (\delta u) / \mathrm{d} y)$. The latter augments (\ref{equ:frozen_turb_variation}) towards a consistent total variation 
\begin{align}
    \delta_\mathrm{u} L \cdot \delta u &= 
    (\nu_\mathrm{eff} + \nu_\mathrm{t}) \frac{\mathrm{d} (\delta u)}{\mathrm{d} y} \bigg \vert_{w} +  \left[(\nu_\mathrm{eff} + \nu_\mathrm{t})\left( \hat{u} \frac{\mathrm{d} (\delta u)}{\mathrm{d} y} - \frac{\mathrm{d} \hat{u}}{\mathrm{d} y} (\delta u) \right) \right]_{w}^{\Delta} 
    &+ \int  \delta u \left[ \frac{\mathrm{d}}{\mathrm{d} y} \left[ (\nu_\mathrm{eff} + \nu_\mathrm{t}) \frac{\mathrm{d} \hat{u} }{\mathrm{d} y} \right] \right] \, \mathrm{d} y
    \qquad \overset{!}{=} 0 \qquad \forall \, \delta u \label{equ:consistent_variation}.
\end{align}
Interestingly, (\ref{equ:consistent_variation}) resembles (\ref{equ:frozen_turb_variation}) by doubling the turbulent contribution.
Hence, the consistent (C) adjoint to (\ref{equ:primal_channel_equation}) reads
\begin{align}
\mathrm{\hat{R}}^\mathrm{\hat{u}, C}: \qquad  \frac{\mathrm{d}}{\mathrm{d} y} \left[ \left( \nu + 2 \nu_\mathrm{t}\right) \frac{\mathrm{d} \hat{u}}{\mathrm{d} y} \right] = 0 \label{equ:consistent_adjoint_equation}.
\end{align}
The asterisk (C) serves to separate the adjoint formulation based on the consistent algebraic turbulence model from the frozen turbulence framework. 
Necessary boundary conditions follow again from the boundary parts in  (\ref{equ:consistent_variation}) and agree with Eqns. (\ref{equ:adjoint_bc_1})-(\ref{equ:adjoint_bc_2}). 

\paragraph{A sensitivity rule} of the objective w.r.t. a general control variable  depends on the definition as well as on the nature of the control. E.g. the relation $\delta u = 0$ (cf. Eqn. (\ref{equ:adjoint_bc_2})) along the channel wall holds as long as the wall is not subjected to control. 
However, if the wall is examined for its optimisation potential, further variational contributions follow from a general shape calculus and are available based on a linear development of the local flow w.r.t. a perturbation in wall normal direction $\delta u = - (\mathrm{d} u / \mathrm{d} y) \delta y$. The latter yields a shape sensitivity derivative expression
\begin{align}
    y=0&: \qquad \delta_\mathrm{y} L \cdot \delta y = 
    \delta_\mathrm{y} j_\mathrm{\Gamma} \cdot \delta y + \nu_\mathrm{eff} \frac{\mathrm{d} \hat{u}}{\mathrm{d} y} \frac{\mathrm{d} u}{\mathrm{d} y} \delta y
    \quad \overset{!}{=} 0 \quad \forall \, \delta y \qquad \rightarrow \qquad 
    \delta_\mathrm{y} j_\mathrm{\Gamma} = - (\nu + \beta \nu_\mathrm{t}) \frac{\mathrm{d} \hat{u}}{\mathrm{d} y} \bigg \vert_{w} \frac{\mathrm{d} u}{\mathrm{d} y} \bigg \vert_{w} \label{equ:shape_derivative}
\end{align}
along the controlled part of the channel boundary and we refer to \cite{soto2004adjoint, soto2004computation, othmer2008continuous, kuhl2019decoupling} for a detailed discussion. The coefficient $\beta = 1$ [$\beta = 2$] accounts for a frozen [consistent] algebraic formulation.
%

%
\section{Law of the Wall}
\label{sec:adjoint_low}
The primal flow description (\ref{equ:primal_channel_equation}) refers to a unidirectional shear flow and assumes a constant near wall stress. 
According to its units, the constant stress $\tau_\mathrm{eff}$ is anticipated to be proportional to the square of a friction velocity $U_{\tau}$, viz. $\tau_\mathrm{eff} := \rho \, U_\mathrm{\tau}^2$. 
The two-layer model assumes a vanishing turbulent stress in the immediate vicinity of the turbulence damping wall ($\mu_\mathrm{t}/\mu \to 0$), frequently labeled as viscous sub-layer, and the opposite behavior beyond a certain wall-normal distance, i.e. inside the logarithmic-layer, where $\mu/\mu_\mathrm{t} \to 0$.
 Using $\nu=\mu/\rho$ and $\nu_t=\mu_t/\rho$, 
Eqn. (\ref{equ:primal_channel_equation}) is usually integrated separately for both limit cases
\begin{alignat}{2}
y < \tilde{y}:& \qquad
U_{\tau}^2 = \nu \frac{\mathrm{d} u}{\mathrm{d} y} 
\qquad &&\rightarrow \qquad
u = \frac{U_{\tau}^2}{\nu} y + C_1 \, , \\
y \ge \tilde{y}:& \qquad
U_{\tau}^2 = (\kappa y)^2 \left| \frac{\mathrm{d} u}{\mathrm{d} y} \right| \frac{\mathrm{d}  u}{\mathrm{d} y} 
\qquad &&\rightarrow \qquad
u = \frac{U_{\tau}}{\kappa} \mathrm{ln}(y)  + C_2 \, , 
\label{eq:low12}
\end{alignat}
where $\tilde{y}$ represents the (theoretical) intersection of the sub- and the logarithmic-layer solution. The use of a no-slip condition along the wall, i.e. at $y = 0$, returns $u_\mathrm{w} = C_1 = 0$. The integration constant $C_2$ is chosen such that the desired transition point is realized and thereby hinges on the choice of $\kappa$. Using non-dimensional parameters based on inner scaling, i.e. $y^+ = U_{\tau} y / \nu$ and $u^+ \coloneqq u / U_{\tau}$, yields a more compact form of the LoW (\ref{eq:low12}),  viz. 
\begin{equation} 
u^+ = \begin{cases}
y^+ & \text{for} \, y^+ < \tilde{y}^+ \\
\frac{1}{\kappa} \ln{(y^+)} + B & \text{for} \, y^+ \geq \tilde{y}^+ \; , 
\end{cases} \label{equ:primal_log_law}
\end{equation}
where the former constant $C_2$ is turned into a non-dimensional constant $B$. Frequently used parameter combinations refer to $\kappa=0.4$ and $B = 5$ to match $\tilde y^+ \approx 11$. In reality the transition from the near-wall to the logarithmic-layer solution spreads over a small region labeled as buffer-layer. 

\paragraph{The adjoint complement} to the LoW (\ref{equ:primal_log_law}) also follows the two-layer ansatz. In line with (\ref{equ:primal_channel_equation}), we first assume an adjoint unidirectional shear flow and a constant adjoint shear stress  
\begin{equation}
\frac{\mathrm{d} \hat{\tau}_\mathrm{eff}}{\mathrm{d} y} = 0 \; , \qquad \to \qquad \hat{\tau}_\mathrm{eff} = \hat{\tau}_\mathrm{l} + \hat{\tau}_\mathrm{t} = (\mu + \beta \mu_{\mathrm{t}}) \frac{\mathrm{d} \hat{u}}{\mathrm{d} y} = \mathrm{const.} =: \rho \, \hat{U}_{\tau}^2 \; \label{equ:adjoint_utau} \; . 
\end{equation}
Equation (\ref{equ:adjoint_utau}) utilizes a coefficient $\beta$ to switch between the  frozen (F; $\beta = 1$) and the consistent (C; $\beta = 2$) algebraic approach. Along the route of  the primal flow, the adjoint stress $\hat{\tau}_\mathrm{eff}$ is assumed to be proportional to the square of an adjoint friction velocity $\hat{U}_{\tau}$.  The two-layer model inherited from the primal flow restricts the effective viscosity of the viscous layer ($\mu_\mathrm{t}/\mu \to 0$) and the log-layer ($\mu/\mu_\mathrm{t} \to 0$). 
%
Analogue to the primal derivation, Eqn. (\ref{equ:adjoint_utau}) is integrated separately for both cases
\begin{alignat}{2}
y \le \tilde{y}:& \qquad
\hat{U}_{\tau}^2 = \nu \frac{\mathrm{d} \hat{u}}{\mathrm{d} y} 
\qquad &&\rightarrow \qquad
\hat{u} = \frac{\hat{U}_{\tau}^2}{\nu} y + \hat{C}_\mathrm{1} \label{equ:adjoint_sublayer} \; ,  \\
y \ge \tilde{y}:& \qquad
\hat{U}_{\tau}^2 = \beta (\kappa y)^2 \frac{\mathrm{d} u}{\mathrm{d} y} \frac{\mathrm{d} \hat{u}}{\mathrm{d} y} 
\qquad &&\rightarrow \qquad
 \hat{u} = \frac{1}{\beta} \left( \frac{\hat{U}_{\tau}}{U_\mathrm{\tau}} \right) \frac{\hat{U}_{\tau}}{\kappa} \mathrm{ln}(y) + \hat{C}_\mathrm{2} \label{equ:adjoint_loglayer} \, . 
\end{alignat}
Note that the primal velocity gradient in the logarithmic regime $\mathrm{d} u / \mathrm{d} y$ was replaced by $ U_\mathrm{\tau} / (\kappa y)$ to solve for the adjoint tangential velocity. Applying a similar velocity normalization, i.e. $\hat{u}^+ \coloneqq \hat{u} / \hat{U}_{\tau}$, yields a compact form of the adjoint LoW similar to  (\ref{equ:primal_log_law}), viz. 
\begin{equation}
\hat{u}^+ = \begin{cases}
y^+ \frac{\hat{U}_\mathrm{\tau}}{U_\mathrm{\tau}} + \frac{\hat{u}_\mathrm{w}}{\hat{U}_\mathrm{\tau}} & \text{for} \, y^+ < \tilde{y}^+ \\
\frac{1}{\beta \, \kappa} \ln{(y^+)} \left( \frac{\hat{U}_\mathrm{\tau}}{U_\mathrm{\tau}} \right) + \hat{B} & \text{for} \, y^+ \ge \tilde{y}^+
\end{cases}. \label{equ:adjoint_log_law}
\end{equation}
Despite a possible shift due to non-intuitive boundary conditions, the adjoint LoW resembles the primal counterpart scaled by the friction velocity ratio $(\hat{U}_\mathrm{\tau} / U_\mathrm{\tau})$  and employs half the logarithmic inclination by the parameter $\beta$ for the consistent approach.

Since the adjoint field quantities are mathematically motivated, 
their adjoint boundary conditions enter the integration constants in Eqn. (\ref{equ:adjoint_sublayer})-(\ref{equ:adjoint_loglayer}). Depending on the objective under investigation, the adjoint velocity potentially experiences a non-zero boundary condition along no-slip walls, hence $\hat{C}_\mathrm{1} = \hat{u}_\mathrm{w}$, e.g. $\hat{C}_\mathrm{1} = -1$ if the shear stress objective from Sec. \ref{sec:derivation} is considered. The piece-wise continuous transition from the sub- towards the logarithmic-layer is ensured by an appropriate value of $\hat{C}_\mathrm{2}$. The latter is reformulated into $\hat{B}$ as an adjoint counterpart of the primal $B$. Using 
\begin{equation}
\tilde{y}^+ \frac{\hat{U}_\mathrm{\tau}}{U_\mathrm{\tau}} + \frac{\hat{u}_\mathrm{w}}{\hat{U}_\mathrm{\tau}}
\, \overset{!}{=} \,
\frac{1}{\beta \, \kappa} \ln{(\tilde{y}^+)} \frac{\hat{U}_\mathrm{\tau}}{U_\mathrm{\tau}} + \hat{B}
\qquad \mathrm{and} \qquad
\tilde{y}^+
\, \overset{!}{=} \,
\frac{1}{\kappa} \ln{(\tilde{y}^+)} + B 
\end{equation}
we conclude that the adjoint $\hat{B}$ follows from the primal $B$, where the latter is augmented by a constant shift in line with the prescribed boundary condition for the adjoint velocity, viz.
\begin{equation}
\hat{B} = \frac{\hat{u}_\mathrm{w}}{\hat{U}_\mathrm{\tau}} + \frac{\hat{U}_\mathrm{\tau}}{U_\mathrm{\tau}} \left[ \frac{B}{\beta} + \tilde{y}^+ \left( 1 - \frac{1}{\beta} \right)\right].
\end{equation}

%
\section{Adjoint Two-Equation Wall Functions}
\label{sec:adjoint_wall_functions}
This section tries to convey the notion that the simple manipulation of adjoint turbulence viscosity also supports more general BVM. We refer to the frequently used baseline $k-\varepsilon$ model \cite{jones1972prediction} as an exemplary turbulence closure of the primal flow equations.
The employed wall boundary conditions are of significance. They refer to standard approaches, used by most engineering finite volume methods, and employ a prescribed shear stress $\tau_w =\tau_\mathrm{eff}$ as well as pressure load on the wall face of the wall adjacent elements to close the primal momentum equations. Zero wall-normal gradients for the turbulent kinetic energy (TKE) $k$ 
and a prescribed near wall value of the  energy dissipation $\varepsilon$, including the assurance of the local turbulence equilibrium $P_\mathrm{k} = \varepsilon$ in the wall adjacent node / cell / element, serve to close the primal turbulence model equations \cite{wilcox1998turbulence}.
The study resembles the investigation already performed in Sec. \ref{sec:derivation} by directly imposing either the primal low-Re or high-Re formulation. 

The algorithmic structure for the low- and the high-Re situation is identical. The only difference 
refers to assigned specific values for the wall shear in line with either of the two solutions (\ref{equ:primal_log_law}), and the near wall value of $\epsilon$ which accommodates to the low- ($\epsilon = 2 \nu k/y^2$) or the high-Re situation ($\epsilon = u_\mathrm{\tau}^3 / (\kappa y) = (\sqrt{c_\mu}k)^{3/2}/(\kappa y)$). As regards the adjoint approach, we only consider the adjoint momentum. The  wall value $\hat u_w$ does -- of course -- not differ for the low- and the high-Re  situation. However, when attention is given to high-Re simulations, the resolution of $\hat u$ in the very near-wall region is deemed computationally expensive and it is  more convenient to follow the same implementation strategy as for the primal flow.
%

In the following, our exemplary objective again refers to the fluid flow induced shear force. 

\paragraph{Employing a low-Re} approach, one frequently imposes 
\begin{align}
        k = 0, \quad \epsilon = 2 \nu \frac{k}{y^2}
        \qquad \text{and thus} \qquad 
        \nu_\mathrm{t} = 0
\end{align}
for the turbulent quantities in the very near-wall regime. This allows for the construction of a Lagrangian, viz.
\begin{align}
        L = \left[ \nu \frac{\mathrm{d} u}{\mathrm{d} y} \right]_\mathrm{w} + \int  \left[ \hat{u} \frac{\mathrm{d}}{\mathrm{d} y} \left[ \nu  \left(\frac{\mathrm{d} u}{\mathrm{d} y}\right) \right] + \hat{k} \left[k \right] + \hat{\epsilon} \left[\epsilon - 2 \nu \frac{k}{y^2} \right] \right] \mathrm{d} y \, . 
        \label{eq:laglow}
\end{align}
The variation of (\ref{eq:laglow})reads
\begin{align}
        \delta L = \left[ \nu \frac{\mathrm{d} (\delta u)}{\mathrm{d} y} \right]_\mathrm{w} + \int  \left[ \hat{u} \frac{\mathrm{d}}{\mathrm{d} y} \left[ \nu  \left(\frac{\mathrm{d} (\delta u)}{\mathrm{d} y}\right) \right] + \hat{k} \left[ \delta k \right] + \hat{\epsilon} \left[\delta \epsilon - 2 \nu \frac{\delta k}{y^2} \right] \right] \mathrm{d} y
\end{align}
and can be rearranged to apply first order optimality conditions, viz.
\begin{align}
        \delta L = \left[ \nu \frac{\mathrm{d} (\delta u)}{\mathrm{d} y} \right]_\mathrm{w} +  \nu \left[ \hat{u} \frac{\mathrm{d} ((\delta u))}{\mathrm{d} y} - \frac{\mathrm{d} \hat{u}}{\mathrm{d} y} ((\delta u)) \right]_\mathrm{w}^\mathrm{\Delta} + \int
        \left[ (\delta u) \,
        \frac{\mathrm{d}}{\mathrm{d} y}  \left[ \nu \frac{\mathrm{d} \hat u}{\mathrm{d} y} \right]  +  \delta k \left[  \hat{k} -\hat{\epsilon} \,  \frac{2 \nu }{y^2} \right] + \delta \epsilon \left[ \hat{\epsilon} \right] \right] \mathrm{d} y  \quad \overset{!}{=} 0 \quad \forall \, (\delta u, \delta k, \delta \epsilon) \label{equ:lagrange_viscous_part_final} \, .
\end{align}
The adjoint low-Re formulation follows from the integral in (\ref{equ:lagrange_viscous_part_final}) and yields
\begin{align}
    \hat{\epsilon} = 0 
    \quad \mathrm{and} \quad 
    \hat{k} =  0
    \qquad \rightarrow \qquad
    \int  \left[ \frac{\mathrm{d}}{\mathrm{d} y}  \left[ \nu  \; \frac{\mathrm{d} \hat u}{\mathrm{d} y}  \right] \right] \mathrm{d} y = 0
\end{align}
and agrees with the observations already documented in the frozen turbulence part of Sec. \ref{sec:derivation}, cf. Eqn. (\ref{equ:frozen_turbadjoint_equation}).
Mind that (\ref{equ:lagrange_viscous_part_final}) is also fulfilled if $\partial k / \partial y$ is employed, hence $\partial (\delta k) / \partial y = 0 \rightarrow \partial \hat{k} / \partial y = 0$.
The boundary conditions for the low-Re formulation follow from the remaining terms in (\ref{equ:lagrange_viscous_part_final}) that can be collected in a compact form and subsequently eliminated, viz.
\begin{alignat}{2}
    y &= \Delta: \qquad \left[ \hat{u} \frac{\mathrm{d} (\delta u)}{\mathrm{d} y} - \frac{\mathrm{d} \hat{u}}{\mathrm{d} y} (\delta u) \right] \qquad \mathrm{with} \qquad \delta \left( \frac{\mathrm{d} u}{\mathrm{d} y} \right) = \frac{\mathrm{d} (\delta u)}{\mathrm{d} y} = 0 \qquad &&\rightarrow \qquad \frac{\mathrm{d} \hat{u}}{\mathrm{d} y} \bigg \vert_\mathrm{\Delta} = 0 \\
    y &= 0: \qquad \left[ (1+\hat{u}) \frac{\mathrm{d} (\delta u)}{\mathrm{d} y}  - \frac{\mathrm{d} \hat{u}}{\mathrm{d} y} (\delta u) \right] \qquad \mathrm{with} \qquad \delta u = 0
    \qquad &&\rightarrow \qquad \hat{u} \big \vert_\mathrm{w} = -1 \; .  \label{equ:low_rn_bounday_conditions}
\end{alignat}
This again confirms the findings of Sec. \ref{sec:derivation} and Eqns. (\ref{equ:adjoint_bc_1})-(\ref{equ:adjoint_bc_2}). 
%
Eqn. (\ref{equ:low_rn_bounday_conditions}) is fulfilled if either
\begin{align}
    \delta u = 0
    \qquad \mathrm{or} \qquad
    \delta u = -\frac{\mathrm{d} u}{\mathrm{d} y} \delta y
    \qquad \rightarrow \qquad \delta_\mathrm{y} j_\mathrm{\Gamma}^\mathrm{LR} = - \nu \frac{\mathrm{d} \hat{u}}{\mathrm{d} y} \bigg \vert_\mathrm{w}  \frac{\mathrm{d} u}{\mathrm{d} y} \bigg \vert_\mathrm{w}
\end{align}
holds that allows for a low-Re (LR) shape derivative expression (cf. Eqn. (\ref{equ:shape_derivative})) if a linear development of the local flow w.r.t. a perturbation in wall normal direction is applied.

\paragraph{Employing a high-Re} $k-\epsilon$ formulation,  
one frequently imposes
\begin{align}
        k = \frac{U_\mathrm{\tau}^2}{\sqrt{C_\mu}} 
        \qquad {\rm and} \qquad 
        \epsilon = \frac{U_\mathrm{\tau}^3}{\kappa y} 
        \qquad \to \qquad 
        \nu_\mathrm{t} = c_\mu \frac{k^2}{\epsilon} = U_\mathrm{\tau} \; \kappa y =  (\kappa y )^2 \frac{\mathrm{d} u}{\mathrm{d} y}.
\end{align}
Hence, a possible Lagrangian, that is valid within the logarithmic layer  [or in the first node / cell / element] from a continuous [discrete] perspective, reads
\begin{align}
        L = \left[ U_\mathrm{\tau} (\kappa y) \frac{\mathrm{d} u}{\mathrm{d} y} \right]_\mathrm{w} + \int  \left[\hat{u} \frac{\mathrm{d}}{\mathrm{d} y} \left[ (\kappa \, y)^2 \left(\frac{\mathrm{d} u}{\mathrm{d} y}\right)^2 \right] + \hat{k} \left[k - \frac{U_\mathrm{\tau}^2}{\sqrt{C_\mu}} \right] + \hat{\epsilon} \left[\epsilon - \frac{U_\mathrm{\tau}^3}{\kappa y} \right] \right] \mathrm{d} y \, . 
\end{align}
Substituting $\epsilon= U_\mathrm{\tau}^3/(\kappa y)= (k \sqrt{c_\mu})^{3/2}/(\kappa y)$ as well as $(\kappa y) \mathrm{d} u/ \mathrm{d} y = U_\mathrm{\tau} = (k \sqrt{C_\mu})^{1/2}$  
we end up with
\begin{align}
        L = \left[ U_\mathrm{\tau} (\kappa y) \frac{\mathrm{d} u}{\mathrm{d} y} \right]_\mathrm{w}  + \int   \left[\hat{u} \frac{\mathrm{d}}{\mathrm{d} y}\left[ (\kappa \, y) \left(\frac{\mathrm{d} u}{\mathrm{d} y} \right) \cdot (k \sqrt{c_\mu})^{1/2} \right] + \hat{k} \left[k - \frac{U_\mathrm{\tau}^2}{\sqrt{C_\mu}} \right] + \hat{\epsilon} \left[\epsilon - \frac{(k \sqrt{C_\mu})^{3/2}}{\kappa y} \right] \right] \mathrm{d} y \, .
\end{align}
A subsequent total variation reads
\begin{align}
        \delta L = \left[ (\delta U_\mathrm{\tau}) (\kappa y) \frac{\mathrm{d} u}{\mathrm{d} y} + (U_\mathrm{\tau} \kappa y)  \frac{\mathrm{d} (\delta u)}{\mathrm{d} y} \right]_\mathrm{w}  &+ \int \bigg[ \hat{u} \frac{\mathrm{d}}{\mathrm{d} y} \left[ (\kappa \, y) \left(\frac{\mathrm{d} (\delta u)}{\mathrm{d} y} \right) \cdot (k \sqrt{C_\mu})^{1/2} +(\kappa \, y) \left(\frac{\mathrm{d}  u}{\mathrm{d} y}\right)
        \delta k  \frac{C_\mu^{1/4}}{2 \sqrt{k}} \right] \nonumber \\
        &\hspace{2cm} + \hat{k} \left[\delta k - 2 \frac{U_\mathrm{\tau}}{\sqrt{C_\mu}} (\delta U_\mathrm{\tau}) \right] + \hat{\epsilon} \left[\delta \epsilon - \delta k \frac{3 \sqrt{k}  C_\mu^{1/4}}{2 \kappa y} \right] \bigg] \mathrm{d} y.
\end{align}
The variations of primal velocity and the TKE are isolated to
\begin{align}
        \delta L &= \bigg[ (\delta U_\mathrm{\tau}) (\kappa y) \frac{\mathrm{d} u}{\mathrm{d} y} + (U_\mathrm{\tau} \kappa y)  \frac{\mathrm{d} (\delta u)}{\mathrm{d} y} \bigg]_\mathrm{w} \nonumber \\
        &+ \bigg[ \hat u \bigg( \frac{\mathrm{d} (\delta u)}{\mathrm{d} y} \; (\kappa y) (k \sqrt{C_\mu})^{1/2} + \delta k \; \frac{\kappa y }{2} \frac{\mathrm{d} u}{\mathrm{d} y} \frac{C_\mu^{1/4}}{\sqrt{k}} \bigg) - (\delta u) \bigg( \frac{\mathrm{d} \hat u}{\mathrm{d} y} \; (\kappa y) (k \sqrt{C_\mu})^{1/2} + 2 \hat k \frac{U_\mathrm{\tau} \kappa y}{\sqrt{C_\mu}}\bigg) \bigg]_\mathrm{w}^\mathrm{\Delta} \nonumber \\
        &+ \int  \bigg[  (\delta u) \, \frac{\mathrm{d}}{\mathrm{d} y}  \bigg[\bigg(\frac{\mathrm{d} \hat u}{\mathrm{d} y}\bigg) (\kappa \, y)  \cdot (k \sqrt{C_\mu})^{1/2} + 2\hat k \, \frac{U_\mathrm{\tau} \kappa y}{\sqrt{C_\mu}} 
        \bigg]  + \delta k \bigg[  \hat{k} -\hat{\epsilon}  \frac{3 \sqrt{k}  C_\mu^{1/4}}{2 \kappa y} - \frac{C_\mu^{1/4}}{2 \sqrt{k}} \, (\kappa \, y) \bigg(\frac{\mathrm{d}  u}{\mathrm{d} y}\bigg) \bigg(\frac{\mathrm{d} \hat u}{\mathrm{d} y}\bigg) \bigg] + \delta \epsilon \left[ \hat{\epsilon} \right] \bigg] \mathrm{d} y \, \label{equ:lagrange_log_part_int}.
\end{align}
Rewriting (\ref{equ:lagrange_log_part_int}) by expressing everything in terms of the primal friction velocity $U_\tau$ yields
\begin{align}
        \delta L &= \bigg[ (\delta U_\mathrm{\tau}) (\kappa y) \frac{\mathrm{d} u}{\mathrm{d} y} + (U_\mathrm{\tau} \kappa y)  \frac{\mathrm{d} (\delta u)}{\mathrm{d} y} \bigg]_\mathrm{w} \nonumber \\
        &+ \bigg[ \hat u \bigg( \frac{\mathrm{d} (\delta u)}{\mathrm{d} y} \; (U_\mathrm{\tau} \kappa y) + \delta k \; \frac{\sqrt{C_\mu} }{2} \bigg) - (\delta u) \bigg( U_\mathrm{\tau} \kappa y \bigg( \frac{\mathrm{d} \hat u}{\mathrm{d} y} \; + \frac{2 \hat k}{\sqrt{C_\mu}} \bigg) \bigg) \bigg]_\mathrm{w}^\mathrm{\Delta} \nonumber \\
        &+ \int  \bigg[ (\delta u) \, \frac{\mathrm{d}}{\mathrm{d} y}  \bigg[ U_\mathrm{\tau} \kappa \, y \bigg( \frac{\mathrm{d} \hat{u}}{\mathrm{d} y} +  \frac{2\hat k}{\sqrt{C_\mu}} \bigg) \bigg]  + \delta k \left[  \hat{k} -\hat{\epsilon} \,  \frac{3 U_\mathrm{\tau} }{2 \kappa y} - \frac{\sqrt{C_\mu}}{2} \,  \bigg(\frac{\mathrm{d} \hat u}{\mathrm{d} y}\bigg) \right] + \delta \epsilon \left[ \hat{\epsilon} \right] \bigg] \mathrm{d} y \quad \overset{!}{=} 0 \quad \forall \, (\delta u, \delta k, \delta \epsilon) \label{equ:lagrange_log_part_final} \; .
\end{align}
Ensuring a vanishing Lagrangian for all possible variations finally yields the adjoint wall functions, viz.
\begin{align}
        \hat{\epsilon} &= 0
        \quad \mathrm{and} \quad
        \hat{k} = \frac{\sqrt{C_\mu}}{2} \frac{\mathrm{d}  \hat u}{\mathrm{d} y} 
        \qquad \rightarrow \qquad
        \int  \left[ \frac{\mathrm{d}}{\mathrm{d} y}  \left[ 2 (U_\mathrm{\tau} \kappa \, y) \; 
        \frac{\mathrm{d} \hat u}{\mathrm{d} y} \right] \right] dy = 0
        \qquad \mathrm{with} \qquad
        U_\mathrm{\tau} \; \kappa y = \nu_\mathrm{t}
        \label{eq:result}
\end{align}
Interestingly, the adjoint dissipation rate is identical zero whereas the adjoint TKE remains as a passive scalar that enters the adjoint shear to form the same expression as in Eqn. (\ref{equ:consistent_variation}) w.r.t. a doubled turbulent viscosity.
The boundary conditions for the high-Re formulation follow from the remaining terms in (\ref{equ:lagrange_log_part_final}) that can be collected in a compact form and subsequently eliminated, viz.
\begin{alignat}{2}
    y &= \mathrm{\Delta} : \qquad \delta k \; \frac{\sqrt{C_\mathrm{\mu}}}{2} \frac{(\kappa y)}{U_\mathrm{\tau}} \frac{\mathrm{d} u}{\mathrm{d} y} \hat{u} + \frac{\mathrm{d} ((\delta u))}{\mathrm{d} y} U_\mathrm{\tau} (\kappa y) \hat{u} - (\delta u) \; \bigg[ 2 (\kappa y) U_\mathrm{\tau} \frac{\mathrm{d} \hat{u}}{\mathrm{d} y} \bigg]
    \quad \overset{!}{=} 0
    \quad \forall \, (\delta k, \delta u) \nonumber && \\
    & \qquad \qquad  \qquad  \qquad \mathrm{with}  \qquad \delta k = 0  \qquad \mathrm{and}  \qquad \frac{ \mathrm{d} ((\delta u))}{\mathrm{d} y} = 0  \qquad &&\rightarrow \qquad
    \frac{ \mathrm{d} \hat{u}}{\mathrm{d} y} \bigg \vert_\mathrm{\Delta} = 0 \\
    y &= 0: \qquad \delta k \bigg[ \frac{\sqrt{C_\mathrm{\mu}}}{2} \frac{(\kappa y)}{U_\mathrm{\tau}} \frac{\mathrm{d} u}{\mathrm{d} y} \bigg[ 1 + \hat{u} \bigg] \bigg] + \frac{\mathrm{d} ((\delta u))}{\mathrm{d} y} \bigg[ U_\mathrm{\tau} (\kappa y) \bigg[ 1 + \hat{u} \bigg] \bigg] - (\delta u) \; \bigg[ 2 (\kappa y) U_\mathrm{\tau} \frac{\mathrm{d} \hat{u}}{\mathrm{d} y} \bigg]
    \qquad &&\overset{!}{=} 0
    \qquad \forall \, (\delta k, \delta u) \nonumber \\
    & \qquad \qquad  \qquad  \qquad \mathrm{with}  \qquad \delta k = 0  \qquad \mathrm{and}  \qquad \delta u = 0  \qquad &&\rightarrow \qquad
    \hat{u} \big \vert_\mathrm{w} = -1 \label{equ:high_rn_bounday_conditions}
\end{alignat}
in line with Eqn. (\ref{equ:shape_derivative}) resulting from the fully continuous derivation in Sec. \ref{sec:derivation}.
Similar to the low-Re formulation, Eqn. (\ref{equ:high_rn_bounday_conditions}) is again fulfilled if either
\begin{align}
    \delta u = 0
    \qquad \mathrm{or} \qquad
    \delta u = -\frac{\mathrm{d} u}{\mathrm{d} y} \delta y
    \qquad \rightarrow \qquad \delta_\mathrm{y} j_\mathrm{\Gamma}^\mathrm{HR} = - 2 (\kappa y) U_\mathrm{\tau} \frac{\mathrm{d} \hat{u}}{\mathrm{d} y} \bigg \vert_\mathrm{w}  \frac{\mathrm{d} u}{\mathrm{d} y} \bigg \vert_\mathrm{w} = -2 \nu_\mathrm{t} \frac{\mathrm{d} \hat{u}}{\mathrm{d} y} \bigg \vert_\mathrm{w} \frac{\mathrm{d} u}{\mathrm{d} y} \bigg \vert_\mathrm{w}
\end{align}
holds that allows for a high-Re (HR) shape derivative expression based on twice the turbulent viscosity (cf. Eqn. (\ref{equ:shape_derivative})).
We thus use stress conditions and prescribe $\hat \tau_w = \hat \tau_\mathrm{eff}$ and $\hat p$ instead of a simple Dirichlet condition $\hat u(y=0) =\hat u_w$, which helps to match with the wall function (\ref{equ:adjoint_log_law}). Mind that, due to the employed objective function, the adjoint flow often features moving walls \cite{kuhl2019decoupling}.

We conclude that there is no need for further adjoint turbulent equations in the range of validity of the adjoint LoW if the above presented wall function based two-equation closure is employed, except a volumetric objective is considered that explicitly depends on the turbulent quantities. 
%
The reason for this is the algebraic scaling of all mean flow and turbulence parameters with the friction velocity $U_\mathrm{\tau}$ within the logarithmic-layer.

%
\section{Verifications}
\label{sec:numerical_results}
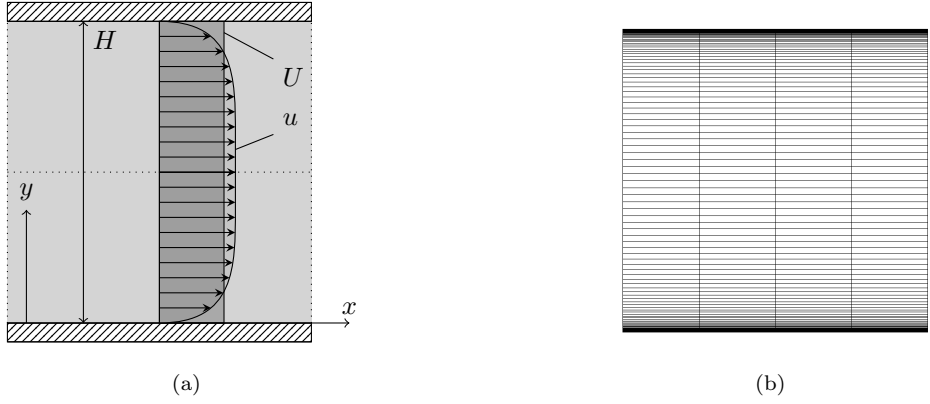
\begin{figure}
	\centering
	\subfigure[]{
		\begin{tikzpicture}
\filldraw[pattern=north east lines, pattern color=black] (0,0) rectangle (4,-0.25);
\filldraw[fill=mycolor_grey!100, draw=none,fill opacity=0.75] (0,0.0) rectangle (4,4);
\draw[thin] (0,0) -- (4,0);
\draw[thin] (0,-0.25) -- (4,-0.25);
\filldraw[pattern=north east lines, pattern color=black] (0,4) rectangle (4,4.25);

\draw[thin] (0,4) -- (4,4);
\draw[thin] (0,4.25) -- (4,4.25);

\draw[dotted] (0,0.0) -- (0,4.0);
\draw[dotted] (4,0.0) -- (4,4.0);

\draw[dotted] (0,2.0) -- (4,2.0);

\draw[thin,->] (0.0,0.0) -- (4.5,0.0) node[anchor=south] {$x$};
\draw[thin,->] (0.25,0.0) -- (0.25,1.5) node[anchor=south] {$y$};

\draw[thin,<->] (1.0,0.0) -- (1.0,4.0);
\draw (1.0,3.75) node[anchor=west] {$H$};

\def \yEnd{2.0};
\def \xpos{2.0};
\def \ypos{0.0};
\def \dx{1.0};
\def \dy{1.5};
\def \ny{10};
\filldraw[fill=mycolor_grey!100,name path=curve] (\xpos,\ypos) .. controls(\xpos+\dx,\ypos) and (\xpos+\dx,\ypos+\dy/2) .. (\xpos+\dx,\ypos+\dy) 
-- (\xpos+\dx,\yEnd)
-- (\xpos,\yEnd)
-- (\xpos,\ypos);
\foreach \y in {1,...,\ny}{
\path[name path=horizontal] (\xpos,\ypos + \yEnd/\ny*\y) -- + (1.0,0);
\draw[-stealth,name intersections={of=curve and horizontal}] (\xpos,\ypos + \yEnd/\ny*\y) -- (intersection-1);
}

\def \yEnd{2.0};
\def \xpos{2.0};
\def \ypos{4.0};
\def \dx{1.0};
\def \dy{-1.5};
\def \ny{10};
\filldraw[fill=mycolor_grey!100,name path=curve] (\xpos,\ypos) .. controls(\xpos+\dx,\ypos) and (\xpos+\dx,\ypos+\dy/2) .. (\xpos+\dx,\ypos+\dy) 
-- (\xpos+\dx,\yEnd)
-- (\xpos,\yEnd)
-- (\xpos,\ypos);
\foreach \y in {1,...,\ny}{
\path[name path=horizontal] (\xpos,\ypos - \yEnd/\ny*\y) -- + (1.0,0);
\draw[-stealth,name intersections={of=curve and horizontal}] (\xpos,\ypos - \yEnd/\ny*\y) -- (intersection-1);
}

\def \yEnd{4.0};
\def \xpos{2.0};
\def \ypos{0.0};
\def \dx{0.85};
\def \dy{1.0};
\def \ny{20};
\filldraw[fill=black,fill opacity=0.2] (\xpos+\dx,0.0) rectangle (\xpos,\yEnd);

\draw[thin,-] (2.85,3.85) -- (3.5,3.5) node[anchor=north west] {$U$};

\draw[thin,-] (3.0,2.3) -- (3.5,2.5) node[anchor=south west] {$u$};

\end{tikzpicture}
	}
	\hspace{2cm}
	\subfigure[]{
		\includegraphics[scale=0.18]{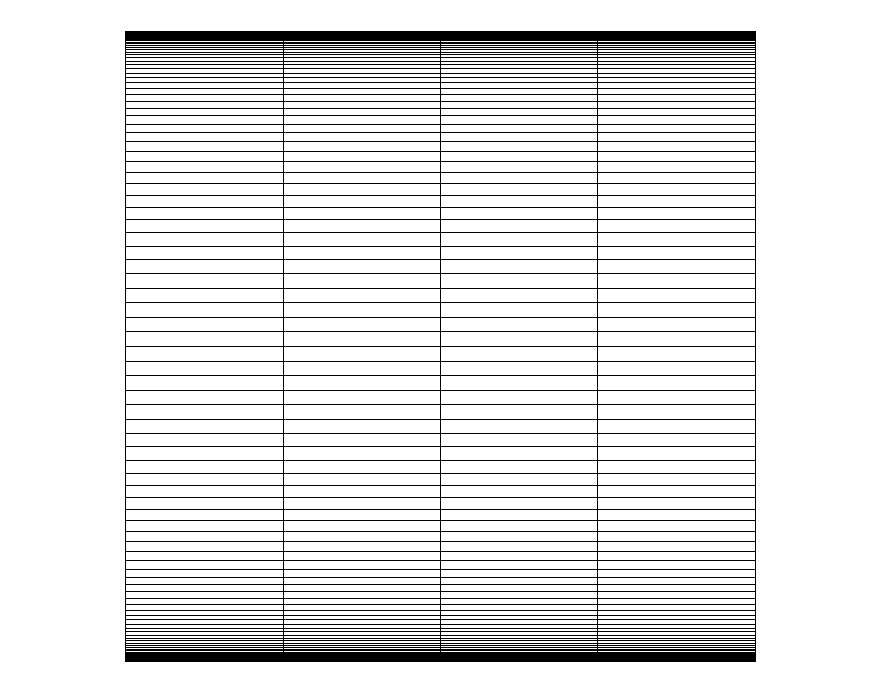}
	}
	\caption{Investigated turbulent channel flow. Sketch of the considered geometry (a) and computational grid (b) for an exemplary Reynolds-number of $\mathrm{Re}_\mathrm{H} = 10^7$.}
	\label{fig:channel_flow}
\end{figure}

The verfication study refers to a 2D turbulent channel flow at Reynolds-numbers between $10^6 \leq \mathrm{Re}_\mathrm{H} = U H / \nu \leq 10^8$ based on the channel height $H$, the bulk velocity $U$ and the kinematic fluid viscosity $\nu$, cf. Fig. \ref{fig:channel_flow}.

\begin{figure}
\centering
\analytiSolutionPictures
\begin{tikzpicture}
\begin{axis}[
 ylabel style={text width=0.25\textwidth,align=center},
 legend style={at={(0.02,0.98)},anchor=north west},
 xlabel={$y^+$ [-]},
 ylabel={$u^+$ [-]},
 xmin=0.2,xmax=10000,
 ymin=0,ymax=30,
 xmode = log,
]

\addplot [line1] table [x expr={\thisrowno{0}},y expr={\thisrowno{1}}] {data/LoW_comparison_lowRe.dat};
\addplot [mark2, only marks] table [x expr={\thisrowno{0}},y expr={\thisrowno{1}}] {data/LoW_comparison_highRe_MixLength.dat};
\addplot [mark3, only marks] table [x expr={\thisrowno{0}},y expr={\thisrowno{1}}] {data/LoW_comparison_highRe_kEps.dat};

\addlegendentry{LoW};
\addlegendentry{Mix. Length};
\addlegendentry{$k - \varepsilon$};
 
\end{axis}
\end{tikzpicture}
\begin{tikzpicture}
\begin{axis}[
 ylabel style={text width=0.25\textwidth,align=center},
 legend style={at={(0.02,0.98)},anchor=north west},
 xlabel={$y^+$ [-]},
 ylabel={$(\nu_\mathrm{t}^\mathrm{m.l.} - \nu_\mathrm{t}^\mathrm{k-\varepsilon})/\nu \cdot 10^{3}$ [-]},
 xmin=50,xmax=10000,
 ymin=-1000,ymax=2000,
 xmode = log,
 ytick={-1000,0,1000,2000,3000},
 yticklabels={-1,0,1,2,3},
]

\addplot [line1] table [x expr={\thisrowno{0}},y expr={\thisrowno{1}}] {data/muT_comparison_ReH_1E+06.dat};
\addplot [line2] table [x expr={\thisrowno{0}},y expr={\thisrowno{1}}] {data/muT_comparison_ReH_1E+07.dat};
\addplot [line3] table [x expr={\thisrowno{0}},y expr={\thisrowno{1}}] {data/muT_comparison_ReH_1E+08.dat};


\addlegendentry{$\mathrm{Re}_\mathrm{H} = 10^6$};
\addlegendentry{$\mathrm{Re}_\mathrm{H} = 10^7$};
\addlegendentry{$\mathrm{Re}_\mathrm{H} = 10^8$};



\end{axis}
\end{tikzpicture}
\begin{tikzpicture}
\begin{axis}[
 ylabel style={text width=0.25\textwidth,align=center},
 xlabel={$y^+$ [-]},
 ylabel={${|{(u^\prime v^\prime)}^+|}^\mathrm{m.l.} - {|{(u^\prime v^\prime)}^+|}^\mathrm{k-\varepsilon}$ [-]},
 xmin=200,xmax=10000,
 ymin=-0.2,ymax=0.8,
 xmode = log,
]

\addplot [line1] table [x expr={\thisrowno{1}},y expr={\thisrowno{13}}] {data/LoW_Compare_UVPlus_ReH_1E+06.dat};
\addplot [line2] table [x expr={\thisrowno{1}},y expr={\thisrowno{13}}] {data/LoW_Compare_UVPlus_ReH_1E+07.dat};
\addplot [line3] table [x expr={\thisrowno{1}},y expr={\thisrowno{13}}] {data/LoW_Compare_UVPlus_ReH_1E+08.dat};

\end{axis}
\end{tikzpicture}
\caption{Comparison of field values for the normalized mean flow (left), turbulent viscosity (center) and 
 Reynolds stresses ${(\overline{u^\prime v^\prime})}^+ = 
(\overline{\left|u^\prime v^\prime\right|}) / U_\mathrm{\tau}^2$ (right) predicted by a mixing-length (open symbols) and a $k - \varepsilon$ (closed symbols) BVM. 
}
\label{fig:bvm_approaches_comparison}
\end{figure}
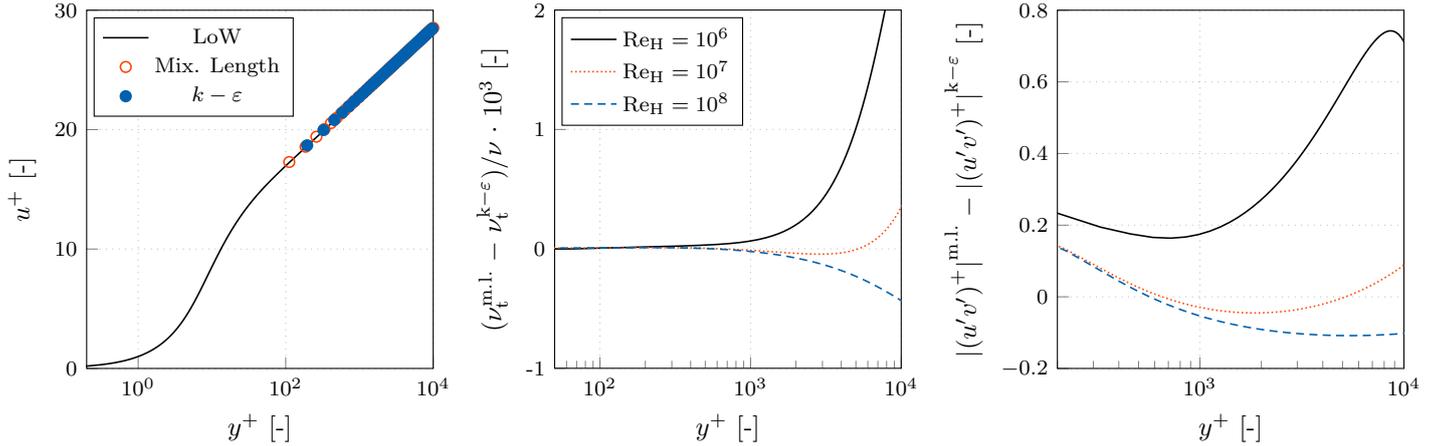

For the sake of clarity, we explicitly state the complete underlying primal and adjoint balance equations used for all verification and application cases. The mean fluid velocity $v_\mathrm{i}$ and pressure $p$ follow from the steady incompressible RANS equations 
\begin{alignat}{2}
R_\mathrm{i}:& \qquad  \rho \, v_\mathrm{k} \frac{\partial v_\mathrm{i}}{\partial x_\mathrm{k}} + \frac{\partial}{\partial x_\mathrm{k}} \left[ \left(p + \frac{2}{3} \rho k \right) \delta_\mathrm{ik} - 2 \left( \mu + \mu_\mathrm{t} \right) S_\mathrm{ik} \right] &&= 0 \label{equ:primal_momentum} \\
Q:& \qquad -\frac{\partial v_\mathrm{k}}{\partial x_\mathrm{k}} &&= 0 \label{equ:primal_mass}.
\end{alignat}
where $S_\mathrm{ik} = 1/2 ( \partial v_\mathrm{i} / \partial x_\mathrm{k} +  \partial v_\mathrm{k} / \partial x_\mathrm{i} )$ and $\delta_\mathrm{ik}$ represent the symmetric strain rate tensor as well as the Kronecker delta respectively.
Mind that we switch the notation compared to the unidirectional setting, viz. $u \coloneqq v_\mathrm{1}$ and $\hat u \coloneqq \hat v_\mathrm{1}$ from now on.
Corresponding boundary conditions are given in \cite{stuck2013adjoint, kroger2018adjoint, kuhl2019decoupling, kuhl2020continuous}.
As already mentioned, wall function expressions often involve singularities at the wall and/or high-order polynomial behaviour beyond the capabilities of the numerical discretization. Thus -- rather than using wall values -- wall function expressions often replace the governing equations in the wall adjacent discrete node / cell / element.
The verification involves two turbulence closures for the primal flow. The low-Re study aims to verify the predictive agreement with the adjoint LoW (\ref{equ:adjoint_log_law}) and therefore employs a mixing-length model supplemented by a van-Driest \cite{vanDriest1956turbulent} damping function $f_\mathrm{vD} = 1 - \mathrm{exp}(-y^+ / A^+)$, i.e. $\nu_\mathrm{t} = (\kappa \, y_w \, f_\mathrm{vD})^2 (\mathrm{d} u / \mathrm{d} y)$. Here $y_w$ represents the normal distance to the nearest wall and $A^+$ was assigned to $A^+ =27$. A standard $k - \epsilon$ model \cite{jones1972prediction} serves as a closure for the high-Re study.
In the absence of volume based objective functional, the adjoint equations to (\ref{equ:primal_momentum})-(\ref{equ:primal_mass}) read:
\begin{alignat}{2}
\hat{R}_\mathrm{i}:& \qquad - \rho v_\mathrm{k} \frac{\partial \hat{v}_\mathrm{i}}{\partial x_\mathrm{k}} + \rho  \hat{v}_\mathrm{k} \frac{\partial v_\mathrm{k}}{\partial x_\mathrm{i}} + \frac{\partial}{\partial x_\mathrm{k}} \left[ \hat{p} \delta_\mathrm{ik} - 2 \left( \mu + \beta \mu_\mathrm{t} \right) \hat{S}_\mathrm{ik} \right] &&= 0 \label{equ:adjoint_momentum} \; \\
\hat{Q}:& \qquad- \frac{\partial \hat{v}_\mathrm{k}}{\partial x_\mathrm{k}} &&= 0 \label{equ:adjoint_mass} \; . 
\end{alignat}
It should be noted that the adjoint equations possibly experience twice the primal turbulent viscosity since $\beta = 2$ [$\beta = 1$] is chosen in the consistent [frozen] case.
Strictly speaking, the suggested approach is only consistent in the immediate wall vicinity. Hence, only the sub-layer and the logarithmic region of the channel flow correspond to a truly consistent adjoint turbulence model. The consistency is lost for the outer layer, while all applications in Sec. \ref{sec:appl} can only refer to a formulation that is deemed to feature an enhanced consistency compared to the frozen turbulence approach.
Mind that shape optimization problems are by definition interested in the primal / adjoint neat wall flow, hence a consistent adjoint formulation is particularly relevant in this region.
Using a two-equation model the consistency is restricted to the momentum equation and assumes the eddy-viscosity distribution to agree with the mixing-length results.
Fig. \ref{fig:bvm_approaches_comparison} 
validates the compliance of both 
approaches for high-Re simulations  
over the normalised wall distance.

A boundary based objective functional is considered that accounts for the fluid flow induced force $J^\mathrm{F}$, viz.
\begin{align}
    J^\mathrm{F} = \int_\mathrm{\Gamma_\mathrm{W}} \left( p \delta_\mathrm{ik} - 2 \mu_\mathrm{eff} S_\mathrm{ik} \right) n_\mathrm{k} r_\mathrm{i} \mathrm{d} \Gamma 
\label{eqn:objectives} \; .
\end{align}
Hence, the local objective reads $j_\mathrm{\Gamma}^\mathrm{F} = ( p \delta_\mathrm{ik} - 2 \mu_\mathrm{eff} S_\mathrm{ik} ) n_\mathrm{k} r_\mathrm{i}$
which in turn enters the adjoint boundary conditions and we refer to \cite{stuck2013adjoint, kuhl2019decoupling} for a detailed overview.
The force objective coincides with the pure shear objective (cf. Sec. \ref{sec:derivation}) augmented by a pressure contribution projected into a certain spatial direction $r_\mathrm{i}$.
After a successfully approximation of the primal and adjoint field equations, a shape sensitivity can be derived along the controlled design wall \cite{stuck2013adjoint, kuhl2019decoupling, kuhl2020adjoint} 
\begin{align}
\delta_\mathrm{u} J 
= \int_{\Gamma_\mathrm{D}} \delta_\mathrm{u} j_\mathrm{\Gamma} \, \mathrm{d} \Gamma_\mathrm{O} 
\qquad \mathrm{with} \qquad 
\delta_\mathrm{u} j_\mathrm{\Gamma} = - \beta \nu_\mathrm{eff}  \frac{\partial v_\mathrm{i}}{\partial x_\mathrm{j}} \frac{\partial \hat{v_\mathrm{i}}}{\partial x_k} n_\mathrm{j} n_\mathrm{k} \label{equ:full_shape_derivative}.
\end{align}

Eqns. (\ref{equ:primal_momentum})-(\ref{equ:adjoint_mass}) are approximated using the Finite-Volume procedure FreSCo+ \cite{rung2009challenges}. Analogue to the use of integration-by-parts in deriving the continuous adjoint equations, summation-by-parts is employed to derive the building blocks of the discrete (dual) adjoint expressions. A detailed derivation of this hybrid adjoint approach can be found in \cite{stuck2013adjoint, kroger2018adjoint, kuhl2020adjoint}. The segregated algorithm uses a cell-centered, collocated storage arrangement for all transport properties. The implicit numerical approximation is second order accurate and supports polyhedral cells. Both, the primal and adjoint pressure-velocity coupling is based on a SIMPLE method and possible parallelization is realized by means of a domain decomposition approach \cite{yakubov2013hybrid, yakubov2015experience}. In terms of the node-based shape optimisation approaches herein, the computational grid is adjusted using a Laplace-Beltrami \cite{stuck2011adjoint, kroger2015cad} [Steklov-Poincar\'e \cite{schulz2016computational, kuhl2020adjoint}] type (surface metric) approach for the external [internal] flows. In all cases, the convective term for primal [adjoint] momentum is approximated using the QUICK [QU(D)ICK] scheme.
%
%
%
Periodic boundary conditions are employed between the inlet and the outlet. A friction condition is used along the top and bottom boundaries in conjunction with low-Re and high-Re grids. The numerical grids consist of $4 \times 250$ finite volumes and the wall normal resolutions reach down to $y^+ = \mathcal{O}(10^{-1})$ for the low-Re cases and $y^+ \approx 50$ for the high-Re cases.
%
%

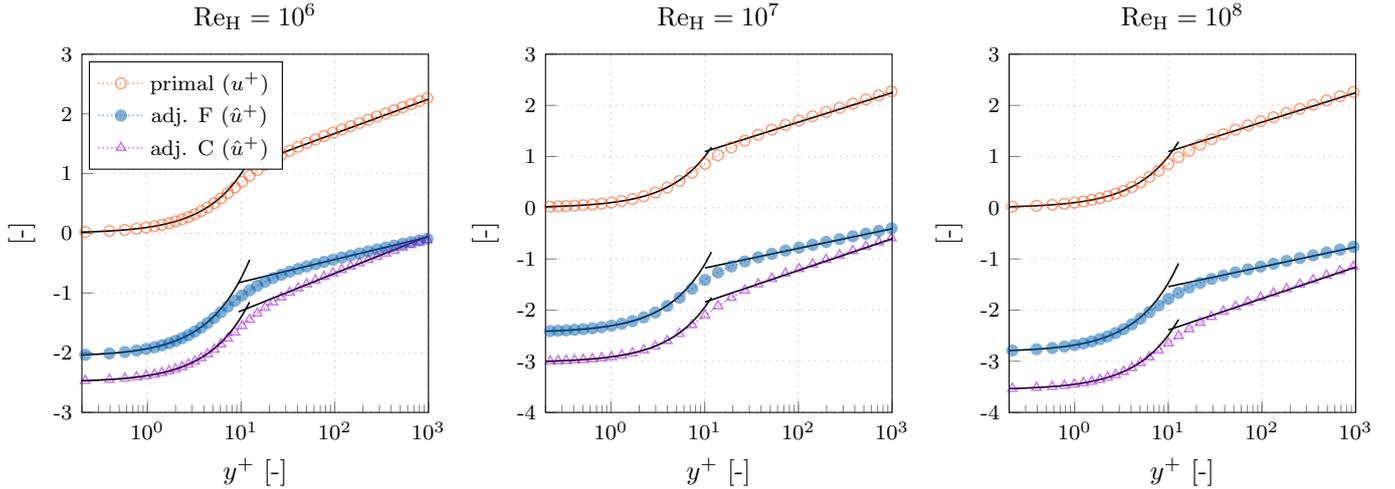
\begin{figure}
\centering
\analytiSolutionPictures
\begin{tikzpicture}
\begin{axis}[
 ylabel style={text width=0.25\textwidth,align=center},
 title={$\mathrm{Re}_\mathrm{H} = 10^6$},
 legend style={at={(0.02,0.98)},anchor=north west},
 xlabel={$y^+$ [-]},
 ylabel={[-]},
 xmin=0.2,xmax=1000,
 ymin=-30,ymax=30,
 xmode = log,
 ytick={-40,-30,-20,-10,0,10,20,30},
 yticklabels={-4,-3,-2,-1,0,1,2,3},
]

\addplot [line2, mark2, each nth point=2, opacity=0.5] table [x=Yplus, y=Uplus] {data/LoW_drag_ReH_1E+06_consistent.dat};
\addplot [line2, mark3, each nth point=2, opacity=0.5] table [x=Yplus, y=adUplus] {data/LoW_drag_ReH_1E+06_consistent.dat};
\addplot [line2, mark4, each nth point=2, opacity=0.5] table [x=Yplus, y=adUplus] {data/LoW_drag_ReH_1E+06_frozenT.dat};

\addplot [line1] table [x=Yplus, y=loglaw1] {data/LoW_drag_ReH_1E+06_consistent.dat};
\addplot [line1] table [x=Yplus, y=loglaw2] {data/LoW_drag_ReH_1E+06_consistent.dat};

\addplot [line1] table [x=Yplus, y=adloglaw1] {data/LoW_drag_ReH_1E+06_consistent.dat};
\addplot [line1] table [x=Yplus, y=adloglaw2] {data/LoW_drag_ReH_1E+06_consistent.dat};

\addplot [line1] table [x=Yplus, y=adloglaw1] {data/LoW_drag_ReH_1E+06_frozenT.dat};
\addplot [line1] table [x=Yplus, y=adloglaw2] {data/LoW_drag_ReH_1E+06_frozenT.dat};

\addlegendentry{primal ($u^+$)};
\addlegendentry{adj. F ($\hat{u}^+$)} ;
\addlegendentry{adj. C ($\hat{u}^+$)};
 
\end{axis}
\end{tikzpicture}
\begin{tikzpicture}
\begin{axis}[
 ylabel style={text width=0.25\textwidth,align=center},
 title={$\mathrm{Re}_\mathrm{H} = 10^7$},
 xlabel={$y^+$ [-]},
 ylabel={[-]},
 xmin=0.2,xmax=1000,
 ymin=-40,ymax=30,
 xmode = log,
 ytick={-40,-30,-20,-10,0,10,20,30},
 yticklabels={-4,-3,-2,-1,0,1,2,3},
]

\addplot [line2, mark2, each nth point=2, opacity=0.5] table [x=Yplus, y=Uplus] {data/LoW_drag_ReH_1E+07_consistent.dat};
\addplot [line2, mark3, each nth point=2, opacity=0.5] table [x=Yplus, y=adUplus] {data/LoW_drag_ReH_1E+07_consistent.dat};
\addplot [line2, mark4, each nth point=2, opacity=0.5] table [x=Yplus, y=adUplus] {data/LoW_drag_ReH_1E+07_frozenT.dat};

\addplot [line1] table [x=Yplus, y=loglaw1] {data/LoW_drag_ReH_1E+07_consistent.dat};
\addplot [line1] table [x=Yplus, y=loglaw2] {data/LoW_drag_ReH_1E+07_consistent.dat};

\addplot [line1] table [x=Yplus, y=adloglaw1] {data/LoW_drag_ReH_1E+07_consistent.dat};
\addplot [line1] table [x=Yplus, y=adloglaw2] {data/LoW_drag_ReH_1E+07_consistent.dat};

\addplot [line1] table [x=Yplus, y=adloglaw1] {data/LoW_drag_ReH_1E+07_frozenT.dat};
\addplot [line1] table [x=Yplus, y=adloglaw2] {data/LoW_drag_ReH_1E+07_frozenT.dat};

\end{axis}
\end{tikzpicture}
\begin{tikzpicture}
\begin{axis}[
 ylabel style={text width=0.25\textwidth,align=center},
 title={$\mathrm{Re}_\mathrm{H} = 10^8$},
 xlabel={$y^+$ [-]},
 ylabel={[-]},
 xmin=0.2,xmax=1000,
 ymin=-40,ymax=30,
 xmode = log,
 ytick={-40,-30,-20,-10,0,10,20,30},
 yticklabels={-4,-3,-2,-1,0,1,2,3},
]
\addplot [line1] table [x=Yplus, y=loglaw1] {data/LoW_drag_ReH_1E+08_consistent.dat};
\addplot [line1] table [x=Yplus, y=loglaw2] {data/LoW_drag_ReH_1E+08_consistent.dat};
\addplot [line2, mark2, each nth point=2, opacity=0.5] table [x=Yplus, y=Uplus] {data/LoW_drag_ReH_1E+08_consistent.dat};

\addplot [line1] table [x=Yplus, y=adloglaw1] {data/LoW_drag_ReH_1E+08_consistent.dat};
\addplot [line1] table [x=Yplus, y=adloglaw2] {data/LoW_drag_ReH_1E+08_consistent.dat};
\addplot [line2, mark3, each nth point=2, opacity=0.5] table [x=Yplus, y=adUplus] {data/LoW_drag_ReH_1E+08_consistent.dat};

\addplot [line1] table [x=Yplus, y=adloglaw1] {data/LoW_drag_ReH_1E+08_frozenT.dat};
\addplot [line1] table [x=Yplus, y=adloglaw2] {data/LoW_drag_ReH_1E+08_frozenT.dat};
\addplot [line2, mark4, each nth point=2, opacity=0.5] table [x=Yplus, y=adUplus] {data/LoW_drag_ReH_1E+08_frozenT.dat};

\end{axis}
\end{tikzpicture}
\caption{Comparison of predicted primal and adjoint velocity profiles using the frozen turbulence (F) as well as the LoW-consistent (C) approach for a turbulent channel flow  at  Reynolds-numbers between $10^6 \leq \mathrm{Re}_\mathrm{H}\leq 10^8$ increasing from left to right (low-Reynolds formulation).
}
\label{fig:loglaw_drag_lowRn}
\end{figure}

\begin{figure}
\centering
\analytiSolutionPictures
\begin{tikzpicture}
\begin{axis}[
 ylabel style={text width=0.25\textwidth,align=center},
 title={$\mathrm{Re}_\mathrm{H} = 10^6$},
 legend style={at={(0.02,0.98)},anchor=north west},
 xlabel={$y^+$ [-]},
 ylabel={[-]},
 xmin=5,xmax=2000,
 ymin=-30,ymax=30,
 xmode = log,
 ytick={-40,-30,-20,-10,0,10,20,30},
 yticklabels={-4,-3,-2,-1,0,1,2,3},
]

\addplot [line2, mark2, each nth point=1, opacity=0.5] table [x=Yplus, y=Uplus] {data/LoW_drag_ReH_1E+06_consistent_highRn.dat};
\addplot [line2, mark3, each nth point=1, opacity=0.5] table [x=Yplus, y=adUplus] {data/LoW_drag_ReH_1E+06_consistent_highRn.dat};
\addplot [line2, mark4, each nth point=1, opacity=0.5] table [x=Yplus, y=adUplus] {data/LoW_drag_ReH_1E+06_frozenT_highRn.dat};

\addplot [line1] table [x=Yplus, y=loglaw1] {data/LoW_drag_ReH_1E+06_consistent.dat};
\addplot [line1] table [x=Yplus, y=loglaw2] {data/LoW_drag_ReH_1E+06_consistent.dat};

\addplot [line1] table [x=Yplus, y=adloglaw1] {data/LoW_drag_ReH_1E+06_consistent.dat};
\addplot [line1] table [x=Yplus, y=adloglaw2] {data/LoW_drag_ReH_1E+06_consistent.dat};

\addplot [line1] table [x=Yplus, y=adloglaw1] {data/LoW_drag_ReH_1E+06_frozenT.dat};
\addplot [line1] table [x=Yplus, y=adloglaw2] {data/LoW_drag_ReH_1E+06_frozenT.dat};

\addlegendentry{primal ($u^+$)};
\addlegendentry{adj. F ($\hat{u}^+$)} ;
\addlegendentry{adj. C ($\hat{u}^+$)};
 
\end{axis}
\end{tikzpicture}
\begin{tikzpicture}
\begin{axis}[
 ylabel style={text width=0.25\textwidth,align=center},
 title={$\mathrm{Re}_\mathrm{H} = 10^7$},
 xlabel={$y^+$ [-]},
 ylabel={[-]},
 xmin=5,xmax=2000,
 ymin=-40,ymax=30,
 xmode = log,
 ytick={-40,-30,-20,-10,0,10,20,30},
 yticklabels={-4,-3,-2,-1,0,1,2,3},
]

\addplot [line2, mark2, each nth point=1, opacity=0.5] table [x=Yplus, y=Uplus] {data/LoW_drag_ReH_1E+07_consistent_highRn.dat};
\addplot [line2, mark3, each nth point=1, opacity=0.5] table [x=Yplus, y=adUplus] {data/LoW_drag_ReH_1E+07_consistent_highRn.dat};
\addplot [line2, mark4, each nth point=1, opacity=0.5] table [x=Yplus, y=adUplus] {data/LoW_drag_ReH_1E+07_frozenT_highRn.dat};

\addplot [line1] table [x=Yplus, y=loglaw1] {data/LoW_drag_ReH_1E+07_consistent.dat};
\addplot [line1] table [x=Yplus, y=loglaw2] {data/LoW_drag_ReH_1E+07_consistent.dat};

\addplot [line1] table [x=Yplus, y=adloglaw1] {data/LoW_drag_ReH_1E+07_consistent.dat};
\addplot [line1] table [x=Yplus, y=adloglaw2] {data/LoW_drag_ReH_1E+07_consistent.dat};

\addplot [line1] table [x=Yplus, y=adloglaw1] {data/LoW_drag_ReH_1E+07_frozenT.dat};
\addplot [line1] table [x=Yplus, y=adloglaw2] {data/LoW_drag_ReH_1E+07_frozenT.dat};

\end{axis}
\end{tikzpicture}
\begin{tikzpicture}
\begin{axis}[
 ylabel style={text width=0.25\textwidth,align=center},
 title={$\mathrm{Re}_\mathrm{H} = 10^8$},
 xlabel={$y^+$ [-]},
 ylabel={[-]},
 xmin=5,xmax=2000,
 ymin=-40,ymax=30,
 xmode = log,
 ytick={-40,-30,-20,-10,0,10,20,30},
 yticklabels={-4,-3,-2,-1,0,1,2,3},
]
\addplot [line2, mark2, each nth point=1, opacity=0.5] table [x=Yplus, y=Uplus] {data/LoW_drag_ReH_1E+08_consistent_highRn.dat};
\addplot [line2, mark3, each nth point=1, opacity=0.5] table [x=Yplus, y=adUplus] {data/LoW_drag_ReH_1E+08_consistent_highRn.dat};
\addplot [line2, mark4, each nth point=1, opacity=0.5] table [x=Yplus, y=adUplus] {data/LoW_drag_ReH_1E+08_frozenT_highRn.dat};

\addplot [line1] table [x=Yplus, y=loglaw1] {data/LoW_drag_ReH_1E+08_consistent.dat};
\addplot [line1] table [x=Yplus, y=loglaw2] {data/LoW_drag_ReH_1E+08_consistent.dat};

\addplot [line1] table [x=Yplus, y=adloglaw1] {data/LoW_drag_ReH_1E+08_consistent.dat};
\addplot [line1] table [x=Yplus, y=adloglaw2] {data/LoW_drag_ReH_1E+08_consistent.dat};

\addplot [line1] table [x=Yplus, y=adloglaw1] {data/LoW_drag_ReH_1E+08_frozenT.dat};
\addplot [line1] table [x=Yplus, y=adloglaw2] {data/LoW_drag_ReH_1E+08_frozenT.dat};

\end{axis}
\end{tikzpicture}
\caption{Comparison of predicted primal and adjoint velocity profiles using the frozen turbulence (F) as well as the LoW-consistent (C) approach for a turbulent channel flow  at  Reynolds-numbers between $10^6 \leq \mathrm{Re}_\mathrm{H}\leq 10^8$ increasing from left to right (high-Reynolds formulation).}
\label{fig:loglaw_drag_highRn}
\end{figure}
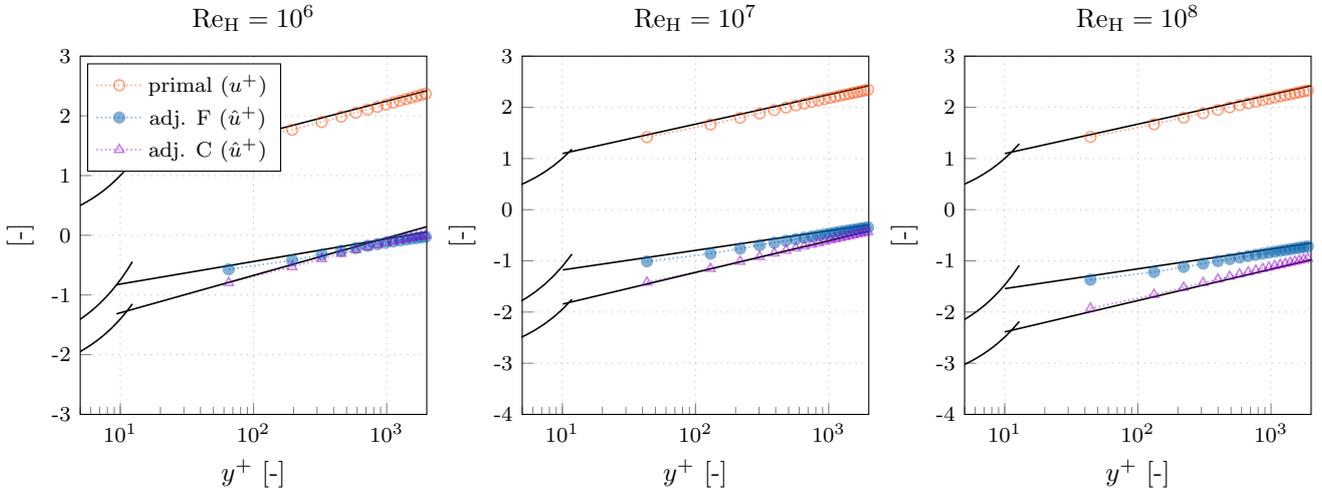

%
%
Figure \ref{fig:loglaw_drag_lowRn} depicts the result of the low-Re studies. 
For all investigated Reynolds numbers, the results are in remarkably fair predictive agreement with the respective LoW (\ref{equ:primal_log_law}) and (\ref{equ:adjoint_log_law}). All results feature a narrow buffer-layer region triggered by the employed van-Driest term.
Figure \ref{fig:loglaw_drag_highRn} depicts the results obtained for the high-Re simulations. It is seen, that the log-layer branch of the two solutions (\ref{equ:primal_log_law}) and (\ref{equ:adjoint_log_law}) is again matched fairly accurate in combination with a $k-\epsilon$ BVM and we conclude, that the adjoint LoW for momentum is compatible with the above suggested approach.
These results encourage us to scrutinize the performance of the simple adjoint algebraic turbulence closure for more complex cases beyond the limits of  unidirectional attached shear flows in the following section.

%
\section{Applications}
\label{sec:appl}
The application part of the manuscript reports the predictive performance of the  algebraic LoW-consistent closure  in more relevant engineering-type flows. Examples included exclusively refer to $k-\varepsilon$ primal flow turbulence modelling in combination with high-Re wall functions. Hence, the wall normal resolutions reach down to $y^+ \approx 50$ for all considered studies.
The focal points of interest  are (a) a comparison of initial shape sensitivities predicted by the adjoint frozen (F) and the LoW-consistent (C) adjoint turbulence closure, and (b) their respective  influence on complete, gradient (steepest descent) based shape optimizations using CAD-free optimization framework.
Initial applications refer to two-dimensional investigations ranging from external to internal flows. The final application refers to the optimization of a three-dimensional ducted geometry.

\subsection{2D External: Pointed Oval}
\label{sec:pointed_oval}
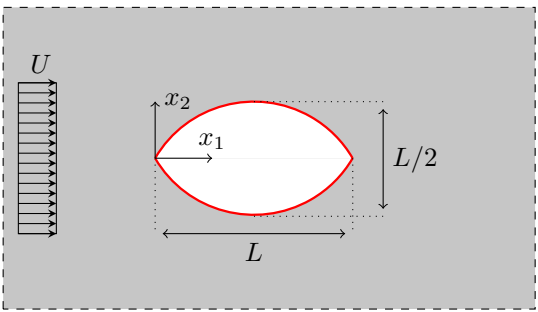
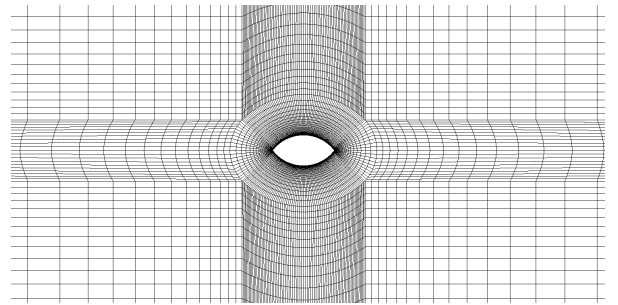
\begin{figure}
	\centering
	\subfigure[]{
		\begin{tikzpicture}

\filldraw[fill=mycolor_grey!100, dashed] (-1,-1) rectangle (6,3);

\filldraw[fill=white, draw=none] (1.0,1.0) arc (-150:-30:1.5);
\filldraw[fill=white, draw=none] (1.0,1.0) arc (150:30:1.5);

\draw[line width=0.3mm,red,-] (1.0,1.0) arc (-150:-30:1.5);
\draw[line width=0.3mm,red,-] (1.0,1.0) arc (150:30:1.5);

\draw[->] (1,1) -- (1.75,1) node [above] {$x_\mathrm{1}$};
\draw[->] (1,1) -- (1,1.75) node [right] {$x_\mathrm{2}$};

\draw[dotted] (1.0,1.0) -- (1.0,0.0);
\draw[dotted] (3.6,1.0) -- (3.6,0.0);
\draw[thin,<->] (1.1,0.0) -- (3.5,0.0);
\draw[] (2.3,-0.0) node[below] {$L$};

\draw[dotted] (2.3,1.75) -- (4.0,1.75);
\draw[dotted] (2.3,0.23) -- (4.0,0.23);
\draw[thin,<->] (4.0,1.65) -- (4.0,0.33);
\draw[] (4,0.99) node[right] {$L/2$};

\def \yEnd{2.0};
\def \ny{15};
\def \xpos{-0.3};
\def \dx{-0.5};
\draw[] (\xpos+\dx,0.0) rectangle (\xpos,\yEnd); 
\foreach \y in {0,...,\ny}{
	\draw[-stealth] (\xpos+\dx,0.0 + \yEnd/\ny*\y) -- (\xpos,0.0+ \yEnd/\ny*\y);
}
\draw[] (-0.5,2.0) node[above] {$U$};

\end{tikzpicture}
	}
	\hspace{2cm}
	\subfigure[]{
		\includegraphics[scale=0.15]{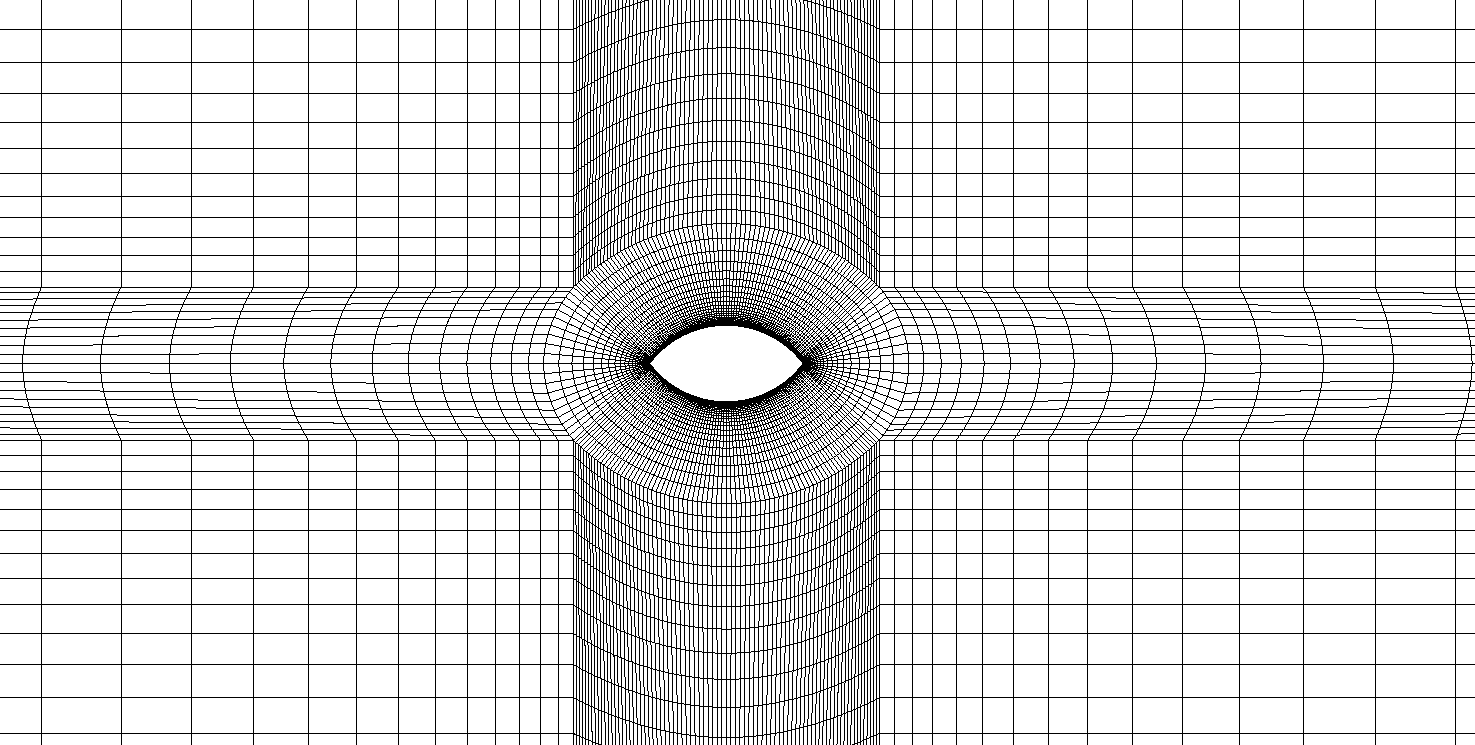}
	}
	\caption{Illustration of the considered geometry (a) and computational grid (b) for the flow around a pointed oval ($\mathrm{Re}_\mathrm{L} = 10^6$. Red lines indicate the design region ).}
	\label{fig:pointed_oval_sketch}
\end{figure}

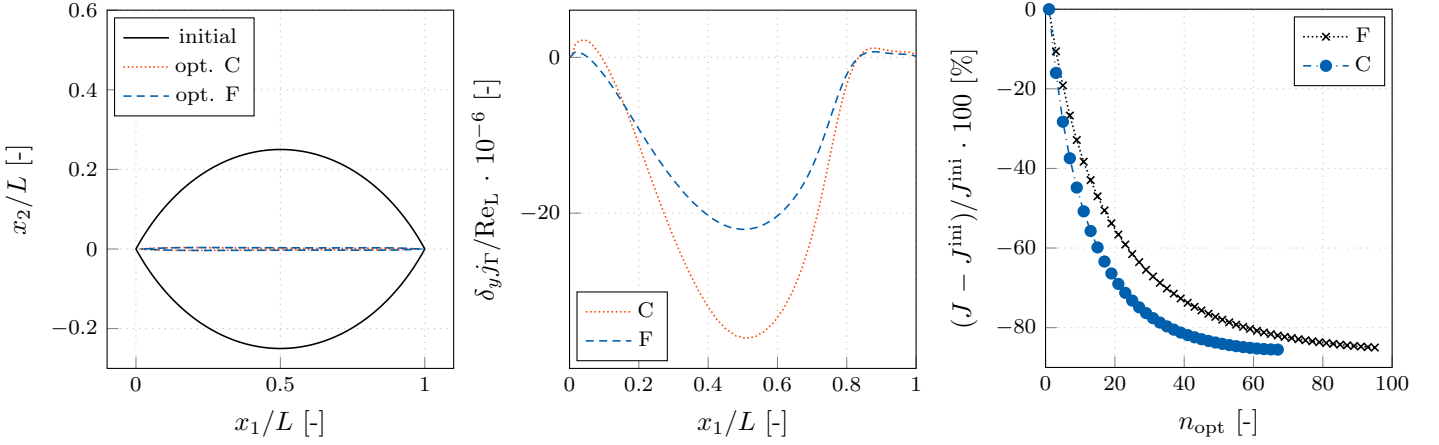
\begin{figure}
\centering
\analytiSolutionPictures
\begin{tikzpicture}
\begin{axis}[
 ylabel style={text width=0.25\textwidth,align=center},
 legend style={at={(0.02,0.98)},anchor=north west},
 xlabel={$x_\mathrm{1}/L$ [-]},
 ylabel={$x_\mathrm{2}/L$ [-]},
 xmin=-0.1,xmax=1.1,
 ymin=-0.3,ymax=0.6,
]

\addplot [line1] table [x expr={\thisrowno{0}},y expr={\thisrowno{1}}] {data/pointed_oval_shapes.dat};
\addplot [line2] table [x expr={\thisrowno{0}},y expr={\thisrowno{1}}] {data/pointed_oval_shapes_without_constraint.dat};
\addplot [line3] table [x expr={\thisrowno{2}},y expr={\thisrowno{3}}] {data/pointed_oval_shapes_without_constraint.dat};

\addlegendentry{initial};
\addlegendentry{opt. C};
\addlegendentry{opt. F}; 
 
\end{axis}

\end{tikzpicture}
\begin{tikzpicture}
\begin{axis}[
 ylabel style={text width=0.25\textwidth,align=center},
 legend style={at={(0.02,0.02)},anchor=south west},
 xlabel={$x_\mathrm{1}/L$ [-]},
 ylabel={$\delta_y j_\mathrm{\Gamma} / \mathrm{Re}_\mathrm{L} \cdot 10^{-6}$ [-]},
 xmin=0.0,xmax=1,
]

\addplot [line2] table [x expr={\thisrowno{0}},y expr={\thisrowno{1}}] {data/pointed_oval_sens.dat};
\addplot [line3] table [x expr={\thisrowno{2}},y expr={\thisrowno{3}}] {data/pointed_oval_sens.dat};
 
\addlegendentry{C};
\addlegendentry{F}; 
 
\end{axis}
\end{tikzpicture}
\begin{tikzpicture}
\begin{axis}[
 ylabel style={text width=0.25\textwidth,align=center},
 legend style={at={(0.98,0.98)},anchor=north east},
 xlabel={$n_{\mathrm{opt}}$ [-]},
 ylabel={$(J-J^{\mathrm{ini}})/J^{\mathrm{ini}} \cdot 100$ [\%]},
 xmin=0,xmax=100,
 ymin=-90,ymax=0,
]

\addplot [line2,mark1,each nth point=2] table [x expr={\thisrowno{0}},y expr={\thisrowno{1}}] {data/pointed_oval_beta_wall_1_beta_field_1_without_constrain.dat};
\addplot [line4,mark3,each nth point=2] table [x expr={\thisrowno{0}},y expr={\thisrowno{1}}] {data/pointed_oval_beta_wall_2_beta_field_2_without_constrain.dat};

\addlegendentry{F};
\addlegendentry{C};

\end{axis}
\end{tikzpicture}
\caption{Initial and optimized shapes (left), initial upper wall shape sensitivities predicted by the frozen (F) and consistent (C) approach, as well as drag objective convergence (right) for the geometrically unconstrained optimization of the flow around a pointed oval at $\mathrm{Re}_\mathrm{L} = 10^6$.
%
}
\label{fig:pointed_oval_results_without_constraints}
\end{figure}

The first test case examines a two dimensional elliptic, pointed oval geometry of length [height] $L$ [$L/2$] under a Reynolds-number of $\mathrm{Re}_\mathrm{L} = U \, L / \nu = 10^6$ where $U$ and $\nu$ refer to the bulk velocity and kinematic viscosity respectively, cf. Fig. \ref{fig:pointed_oval_sketch} (a) and \ref{fig:pointed_oval_results_without_constraints} (left). The investigated oval employs a height $h$ to length ratio of $h/L = 1/2$. 
The structured numerical grid consists of 11\,600 control volumes and the obstacle is discretized with 200 surface elements as depicted in Fig. \ref{fig:pointed_oval_sketch} (b).
A homogeneous velocity is imposed along the inlet, a zero pressure value is prescribed at the outlet and slip-walls are employed along the top and bottom boundary.
The obstacle is optimized w.r.t. the total resistance $J^\mathrm{F}$ (cf. Eqn. (\ref{eqn:objectives})). In line with the habitat of the objective (\ref{eqn:objectives}), the adjoint velocity reads $\hat{v}_\mathrm{i} = -r_\mathrm{i} = -\delta_\mathrm{1i}$  along the design surface.

The initial shape sensitivities along the upper side resulting from both employed adjoint formulations (F vs. C) are shown in Fig. \ref{fig:pointed_oval_results_without_constraints} (center). Notable quantitative differences are observed in the maximum absolute sensitivity. Moreover, qualitative differences occur due to the deviating signs in the vicinity of the leading and trailing edge. 
While the former should primarily result in an accelerated optimization, the latter points to possibly different optimal solutions.
For this reason, two optimizations were performed, i.e. one for a convex problem, that should exclusively reveal convergence speed differences, and one for a non-convex problem.

The first study uses an identical step size in combination with a Laplace-Beltrami surface metric, which extracts the inherently smooth shape gradient out of the possibly rough shape derivative, viz. $g - h^{2} \Delta_\mathrm{\Gamma} g = s$ where $g$ and $\Delta_\mathrm{\Gamma} = \partial^2 / \partial x_\mathrm{k}^{2} - \partial^2 / \partial n^{2}$ represent the shape gradient as well as the Laplace-Beltrami operator respectively \cite{stuck2011adjoint,kroger2015cad}. The utilized step size was chosen to ensure a maximum first displacement $d$ of $ d / L = 1/1000$ for the consistent optimization. 
The mathematically convex problem should physically converge to a flat plate boundary layer flow. The convergence of the drag objective is documented in Fig. \ref{fig:pointed_oval_results_without_constraints} (right) where both strategies yield almost the same optimal value that drops by approximately 85\%. However, the LoW-consistent approach converges approximately 30\% faster compared to the frozen turbulence approach.
The optimized shapes are depicted in Fig. \ref{fig:pointed_oval_results_without_constraints} (left) and the deviation of their optimal drag value is below 2\% w.r.t. the non-dimensional drag coefficient of a turbulent flat plate boundary layer, e.g. $c_\mathrm{d} \approx 0.074 \, \mathrm{Re}_\mathrm{L}^{(-1/5)} = 4.7 \cdot 10^{-3}$, cf. \cite{hucho2002aerodynamik}.

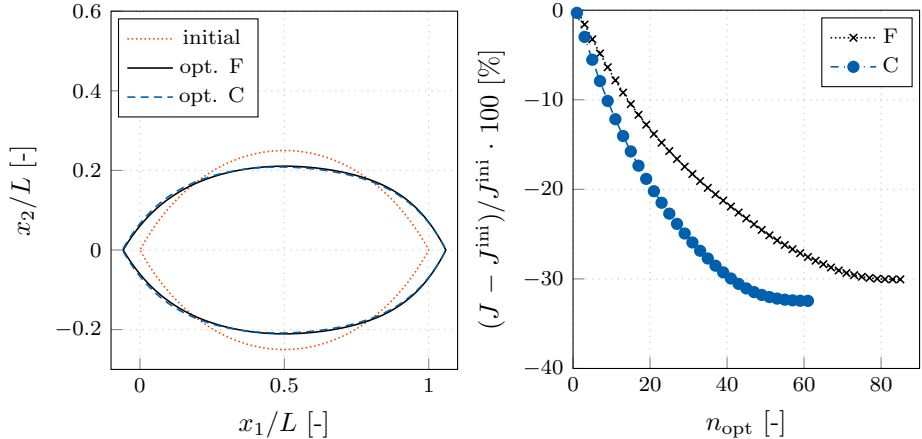
\begin{figure}
\centering
\analytiSolutionPictures
\begin{tikzpicture}
\begin{axis}[
 ylabel style={text width=0.25\textwidth,align=center},
 legend style={at={(0.02,0.98)},anchor=north west},
 xlabel={$x_\mathrm{1}/L$ [-]},
 ylabel={$x_\mathrm{2}/L$ [-]},
 xmin=-0.1,xmax=1.1,
 ymin=-0.3,ymax=0.6,
]

\addplot [line2] table [x expr={\thisrowno{0}},y expr={\thisrowno{1}}] {data/pointed_oval_shapes.dat};
\addplot [line1] table [x expr={\thisrowno{4}},y expr={\thisrowno{5}}] {data/pointed_oval_shapes.dat};
\addplot [line3] table [x expr={\thisrowno{2}},y expr={\thisrowno{3}}] {data/pointed_oval_shapes.dat};

\addlegendentry{initial};
\addlegendentry{opt. F};
\addlegendentry{opt. C}; 
 
\end{axis}

\end{tikzpicture}
\begin{tikzpicture}
\begin{axis}[
 ylabel style={text width=0.25\textwidth,align=center},
 legend style={at={(0.98,0.98)},anchor=north east},
 xlabel={$n_{\mathrm{opt}}$ [-]},
 ylabel={$(J-J^{\mathrm{ini}})/J^{\mathrm{ini}} \cdot 100$ [\%]},
 xmin=0,xmax=90,
 ymin=-40,ymax=0,
]

\addplot [line2,mark1,each nth point=2] table [x expr={\thisrowno{0}},y expr={\thisrowno{1}}] {data/pointed_oval_beta_wall_1_beta_field_1.dat};
\addplot [line4,mark3,each nth point=2] table [x expr={\thisrowno{0}},y expr={\thisrowno{1}}] {data/pointed_oval_beta_wall_2_beta_field_2.dat};

\addlegendentry{F};
\addlegendentry{C};

\end{axis}
\end{tikzpicture}
%
%
%
%
\caption{Initial and optimized shapes (left) as well as drag objective convergence (right) predicted by the frozen (F) and the LoW-consistent (C) approaches for the volume conserving optimization of the flow around a pointed oval at $\mathrm{Re}_\mathrm{L} = 10^6$.}
\label{fig:pointed_oval_results}
\end{figure}

Subsequently, an additional optimization study was carried out, whereby the sensitivity is modified in such that the flow displacement of the initial shape is conserved using a projection method, viz. $s \to s - \int s_\mathrm{i} n_\mathrm{i} \mathrm{d} \Gamma / \int 1 \mathrm{d} \Gamma$.
Analogous to the previous optimization, the same constant step size was specified for both optimizations, which was chosen to ensure a maximum first displacement of $d / L = 1 / 1000$ for the consistent optimization. The convergence of the objective function is documented in the right graph of Fig. \ref{fig:pointed_oval_results}. Again, the LoW-consistent approach converges almost 30\% faster, while absolute [relative] improvements of $\approx$3\% [$\approx$10\%]  are observed for the resistance reduction compared to the frozen turbulence approach. 
The profit follows mainly from the slightly more bulbous [slimmer] front [rear] region (cf. Fig. \ref{fig:pointed_oval_results}).

An interesting aspect follows from a comparison of the optimal resistance reduction observed with different adjoint algebraic turbulence models. Lifting the ratio between the primal and the adjoint eddy-viscosity from the LoW-consistent value of 2 to similar values -- e.g. 3, 4 or 5 -- inside the field does marginally change the computed optimum and the related behaviour is equivocal.  More drastic changes are detrimental to the accuracy and robustness. Hence, we would recommend to retain the LoW-consistent value.

\subsection{2D T-Junction Flow}
\label{sec:tjunction}
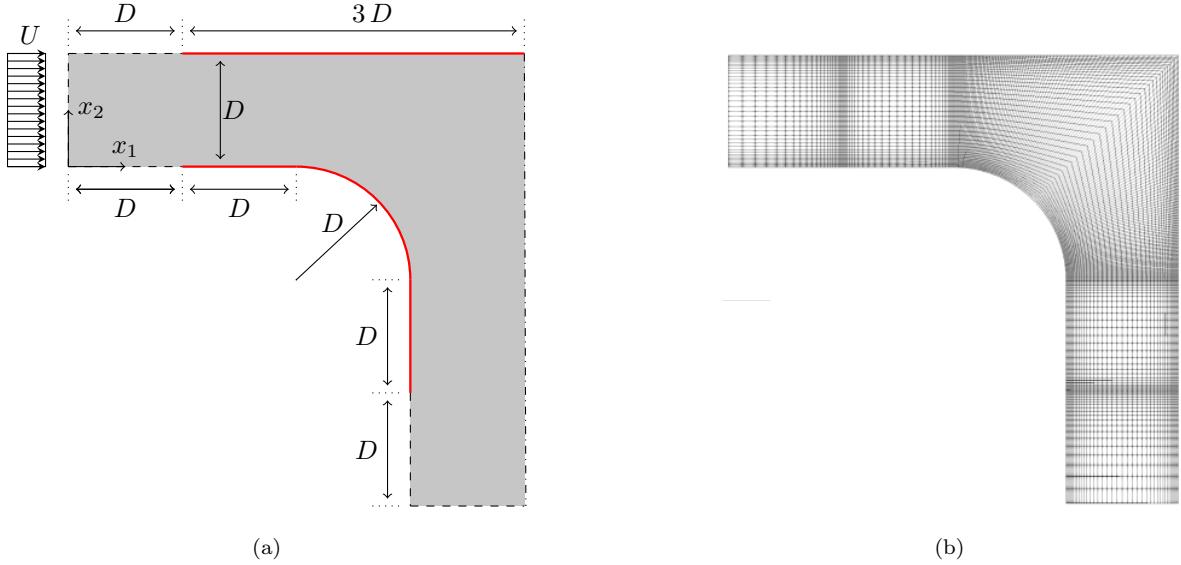
\begin{figure}
	\centering
	\subfigure[]{
		\begin{tikzpicture}

\filldraw[fill=mycolor_grey!100, draw=none] (0,0) rectangle (6,1.5);
\filldraw[fill=mycolor_grey!100, draw=none] (4.5,0.75) rectangle (6,-4.5);

\filldraw[fill=mycolor_grey!100, draw=none] (3,0.75) rectangle (5,-1.5);
\fill [white] (3,-1.5) circle (1.5);

\draw[dashed] (0,0) -- (0,1.5);
\draw[dashed] (0,0) -- (1.5,0);
\draw[dashed] (0,1.5) -- (1.5,1.5);
\draw[dashed] (4.5,-3) -- (4.5,-4.5);
\draw[dashed] (4.5,-4.5) -- (6,-4.5);

\draw[dotted] (0,1.6) -- (0,2.0);
\draw[dotted] (1.5,1.6) -- (1.5,2.0);
\draw[dotted] (6,1.6) -- (6,2.0);

\draw[dotted] (0,-0.1) -- (0,-0.5);
\draw[dotted] (1.5,-0.1) -- (1.5,-0.5);
\draw[dotted] (3,-0.1) -- (3,-0.5);

\draw[dotted] (4.0,-1.5) -- (4.4,-1.5);
\draw[dotted] (4.0,-3.0) -- (4.4,-3.0);
\draw[dotted] (4.0,-4.5) -- (4.4,-4.5);

\draw[dashdotted] (6.0,1.5) -- (6.02,-4.5);

\draw[thin,<->] (0.1,1.8) -- (1.4,1.8);
\draw[thin,<->] (1.6,1.8) -- (5.9,1.8);
\draw[thin,<->] (0.1,-0.3) -- (1.4,-0.3);
\draw[thin,<->] (0.1,-0.3) -- (1.4,-0.3);
\draw[thin,<->] (1.6,-0.3) -- (2.9,-0.3);
\draw[thin,<->] (4.2,-1.6) -- (4.2,-2.9);
\draw[thin,<->] (4.2,-3.1) -- (4.2,-4.4);
\draw[thin,<->] (2.0,0.1) -- (2.0,1.4);
\fill [black] (3,-1.5) circle (0.01);
\draw[thin,->] (3,-1.5) -- (4.0606,-0.50606);

\draw[line width=0.3mm,red,-] (1.5,1.5) -- (6,1.5);
\draw[line width=0.3mm,red,-] (1.5,0) -- (3,0);
\draw[line width=0.3mm,red,-] (4.5,-1.5) -- (4.5,-3);
\draw[line width=0.3mm,red,-] (4.5,-1.5) arc (0:90:1.5);

\draw[] (0.75,1.8) node[above] {$D$};
\draw[] (4,1.8) node[above] {$3 \, D$};
\draw[] (1.9,0.75) node[right] {$D$};
\draw[] (0.75,-0.3) node[below] {$D$};
\draw[] (2.25,-0.3) node[below] {$D$};
\draw[] (4.2,-2.25) node[left] {$D$};
\draw[] (4.2,-3.75) node[left] {$D$};
\draw[] (3.75,-0.75) node[left] {$D$};

\def \yEnd{1.5}
\def \ny{15};
\def \xpos{-0.3};
\def \dx{-0.5};
\draw[] (\xpos+\dx,0.0) rectangle (\xpos,\yEnd); 
\foreach \y in {0,...,\ny}{
	\draw[-stealth] (\xpos+\dx,0.0 + \yEnd/\ny*\y) -- (\xpos,0.0+ \yEnd/\ny*\y);
}
\draw[] (-0.5,1.5) node[above] {$U$};

\draw[->] (0,0) -- (0.75,0) node [above] {$x_\mathrm{1}$};
\draw[->] (0,0) -- (0,0.75) node [right] {$x_\mathrm{2}$};






\end{tikzpicture}
	}
	\hspace{2cm}
	\subfigure[]{
		\includegraphics[scale=0.6]{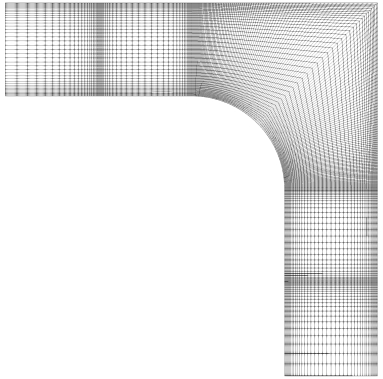}
	}
	\caption{Sketch of the considered symmetric geometry (a) and computational grid (b) for the turbulent T-junction study at  $\mathrm{Re}_\mathrm{D} = 5 \cdot 10^4$. Red lines indicate the design region.}
	\label{fig:tjunction_sketch}
\end{figure}

The second test case examines a two-dimensional T-junction at a bulk Reynolds-number of $\mathrm{Re}_\mathrm{D} = U \, D / \nu = 5 \cdot 10^4$ where $U$, $D$ and $\nu$ refer to the bulk velocity, inlet diameter as well as the kinematic viscosity respectively, cf. Fig. \ref{fig:tjunction_sketch} (a).
The structured numerical grid models half of the geometry and consists of 10\,000 control volumes. The upper/flat [inner/curved] boundary is free for design and discretized with 105 [210] surface elements as depicted in Fig. \ref{fig:tjunction_sketch} (b). The grid is refined towards the transition between fixed and designed wall.
Along the inlet, a homogeneous velocity is imposed together with turbulent quantities that follow from the empirical relation
\begin{align}
    k = \frac{3}{2} \mathrm{Tu}^2 U^2 \, ,
    \quad
    \epsilon = \sqrt{\frac{3}{2}} \frac{\mathrm{Tu} \, U}{0.07 D} C_\mathrm{\mu}^{3/4} k
    \qquad \mathrm{and} \qquad
    \mathrm{Tu} = 0.16 \, \mathrm{Re}_\mathrm{D}^{-1/8} \; . \label{equ:turbulence_init}
\end{align}
A zero pressure value is prescribed at the outlet.
The ducted geometry is optimized w.r.t. the total power loss $J^\mathrm{P}$ 
\begin{align}
    J^\mathrm{P} = - \int_\mathrm{\Gamma_\mathrm{in,out}} n_\mathrm{k} v_\mathrm{k} \left( p + \frac{\rho}{2} v_\mathrm{i}^2 \right) \mathrm{d} \Gamma 
\label{eqn:objectives2} \; .
\end{align}
 which yields $j_\mathrm{\Gamma}^\mathrm{P} = -n_\mathrm{k} v_\mathrm{k} ( p + \frac{\rho}{2} v_\mathrm{i}^2 )$.
In line with the habitat of the objective (\ref{eqn:objectives2}), the adjoint pressure is prescribed to ensure $\hat{p} n_\mathrm{i} = \rho v_\mathrm{k} n_\mathrm{k} \hat{v}_\mathrm{i} + \mu_\mathrm{eff} (\partial \hat{v}_\mathrm{i} / \partial x_\mathrm{k}) n_\mathrm{k} - 0.5 \rho v_\mathrm{k}^2 n_\mathrm{i} - \rho v_\mathrm{k} n_\mathrm{k} v_\mathrm{i}$ along the outlet whereas the adjoint velocity is defined as $\hat{v}_\mathrm{i} = v_\mathrm{i}$ at the inlet, cf. \cite{stuck2013adjoint}.

\begin{figure}
\centering
\analytiSolutionPictures
\begin{tikzpicture}
\begin{axis}[
 ylabel style={text width=0.25\textwidth,align=center},
 legend style={at={(0.02,0.98)},anchor=north west},
 xlabel={$x_\mathrm{1}/D$ [-]},
 ylabel={$\delta_y j_\mathrm{\Gamma} / \mathrm{Re}_\mathrm{H} \cdot 10^{-5}$ [-]},
 xmin=0.0,xmax=4,
 ymin=-30,ymax=5,
]

\addplot [line2] table [x expr={\thisrowno{2}},y expr={\thisrowno{3}}] {data/tjunction_outer_ring_consistent.dat};
\addplot [line3] table [x expr={\thisrowno{2}},y expr={\thisrowno{3}}] {data/tjunction_outer_ring_frozen.dat};

\addlegendentry{C};
\addlegendentry{F} ; 
 
\end{axis}

\end{tikzpicture}
\begin{tikzpicture}
\begin{axis}[
 ylabel style={text width=0.25\textwidth,align=center},
 legend style={at={(0.02,0.98)},anchor=north west},
 xlabel={$(x_1 - x_2)/D$ [-]},
 ylabel={$\delta_y j_\mathrm{\Gamma} / \mathrm{Re}_\mathrm{H} \cdot 10^{-5}$ [-]},
 xmin=0.0,xmax=6,
 ymin=-11,ymax=1,
]

\addplot [line2] table [x expr={\thisrowno{2}},y expr={\thisrowno{3}}] {data/tjunction_inner_ring_consistent.dat};
\addplot [line3] table [x expr={\thisrowno{2}},y expr={\thisrowno{3}}] {data/tjunction_inner_ring_frozen.dat};
 
\end{axis}
\end{tikzpicture}
\begin{tikzpicture}
\begin{axis}[
 ylabel style={text width=0.25\textwidth,align=center},
 legend style={at={(0.98,0.98)},anchor=north east},
 xlabel={$n_{\mathrm{opt}}$ [-]},
 ylabel={$(J-J^{\mathrm{ini}})/J^{\mathrm{ini}} \cdot 100$ [\%]},
 xmin=0,xmax=30,
 ymin=-15,ymax=0,
]

\addplot [line2,mark1] table [x expr={\thisrowno{0}},y expr={\thisrowno{1}}] {data/tjunction_betaWall_1_betaField_1.dat};
\addplot [line4,mark3] table [x expr={\thisrowno{0}},y expr={\thisrowno{1}}] {data/tjunction_betaWall_2_betaField_2.dat};

\addlegendentry{F};
\addlegendentry{C};


\end{axis}
\end{tikzpicture}
\caption{Local shape derivatives w.r.t. a power loss objective predicted by the LoW-consistent (C) and frozen (F) turbulence approach along the upper/flat (left) and the inner/curved boundary (center) of a turbulent T-junction flow ($\mathrm{Re}_\mathrm{D} = 5 \cdot 10^4$). The right graph  documents the convergence of objective functional over the optimization cycles.}
\label{fig:tjunction_sens_and_conv}
\end{figure}
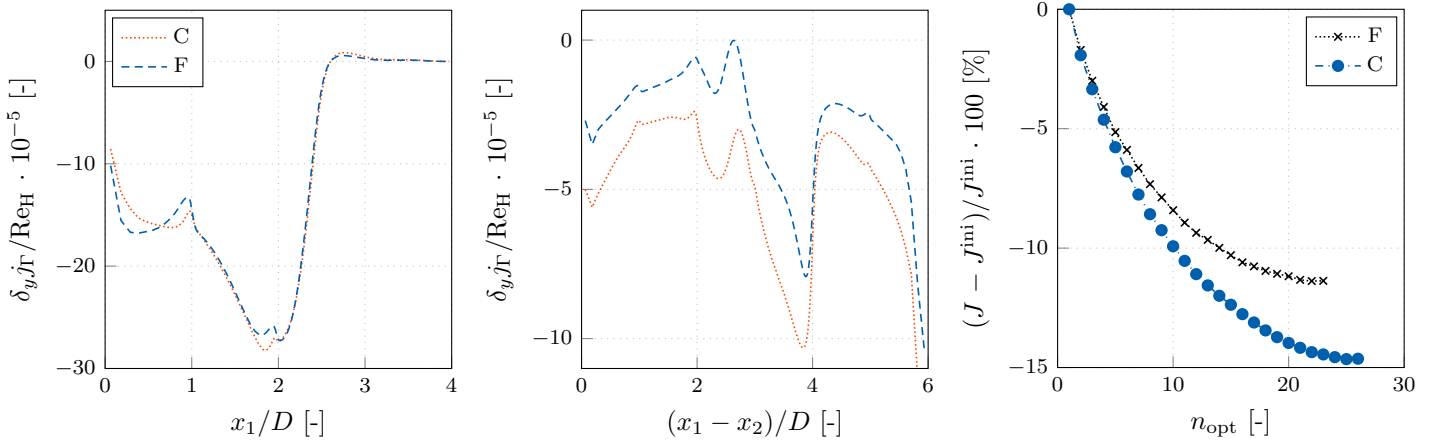

\begin{figure}
\centering
\input{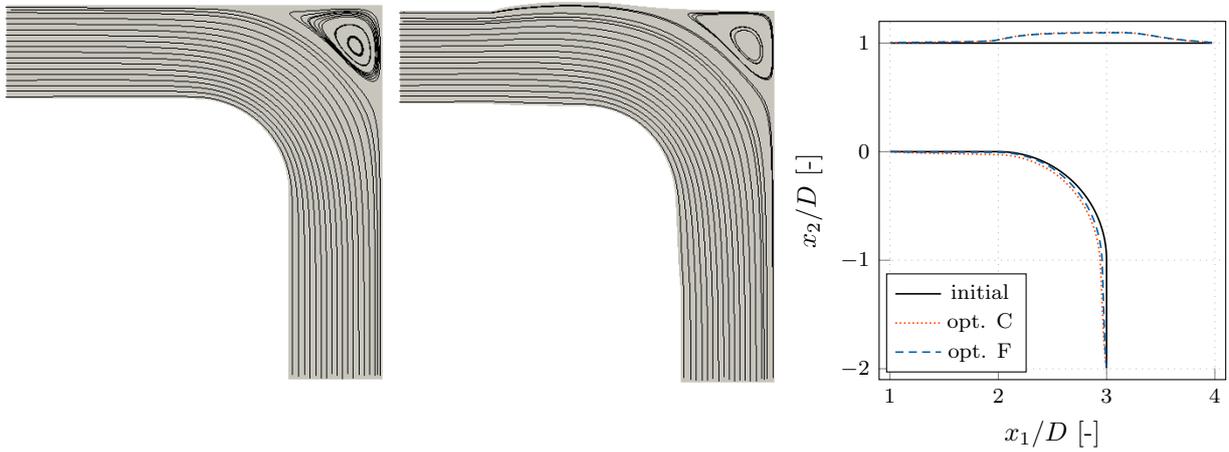}
\caption{Comparison of streamlines for the initial (left) and  optimized T-junction geometry which was returned by the LoW-consistent (C) approach (center; $\mathrm{Re}_\mathrm{D} = 5 \cdot 10^4$). The (right) graph displays a detailed comparison of the design changes.}
\label{fig:tjunction_shapes}
\end{figure}

The initial flow field is depicted in Fig. \ref{fig:tjunction_shapes} (left) where the re-circulation zone is deemed to be responsible for a large portion of the total power loss.
The resulting initial shape sensitivity along the upper/flat [inner/curved] boundary is depicted in Fig. \ref{fig:tjunction_sens_and_conv} (left) [(center)] for both proposed adjoint formulations. Basically, the sensitivities along the two design walls appear affine to each other but a noticeable increase in the sensitivity magnitude arises along the curved design region. The latter is pronounced in the direction of the narrowed region ($(x_\mathrm{1} - x_\mathrm{2})/D \approx 4$) immediately after the bend.
Its influence on a gradient based optimization process is documented in Fig. \ref{fig:tjunction_sens_and_conv} (right). Assuming an equal step size for both optimizations --which follows from the LoW-consistent approach with an initial maximum displacement of $d/D = 1/1000$-- the consistent approach finds a minimum that is absolutely [relatively] $\approx$3.3\% [$\approx$22.2\%] smaller compared to the optimal shape w.r.t. the frozen turbulence approach.
The optimal shape of the consistent approach is depicted in Fig. \ref{fig:tjunction_shapes} (center) and (right). In line with the absolute sensitivity values, cf. Fig. \ref{fig:tjunction_sens_and_conv}, the modification of the initially flat part appears to be pronounced, which finally returns a visible reduction of the re-circulation.

\subsection{3D Double Bent Pipe Flow}
\label{sec:double_bent_pipe}
\begin{figure}
\centering
\input{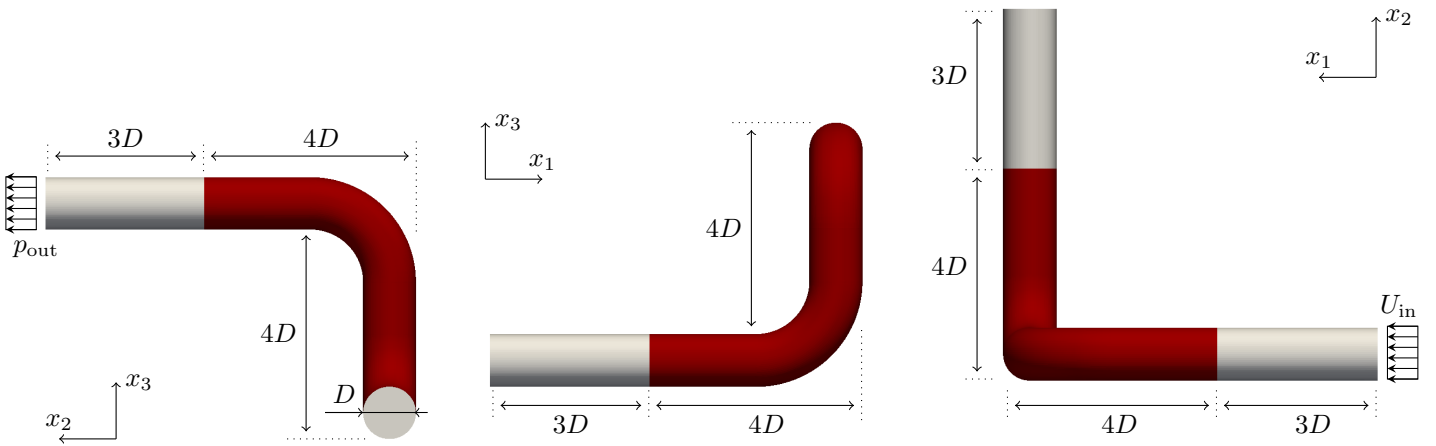}
\caption{Divers views on the initial geometry of the turbulent double bent pipe case  at $\mathrm{Re}_\mathrm{D} = 10^6$. Red areas indicate the design region.}
\label{fig:double_bent_pipe_1}
\end{figure}

\begin{figure}
\centering
\subfigure[]{
\centering
\includegraphics[scale=0.15]{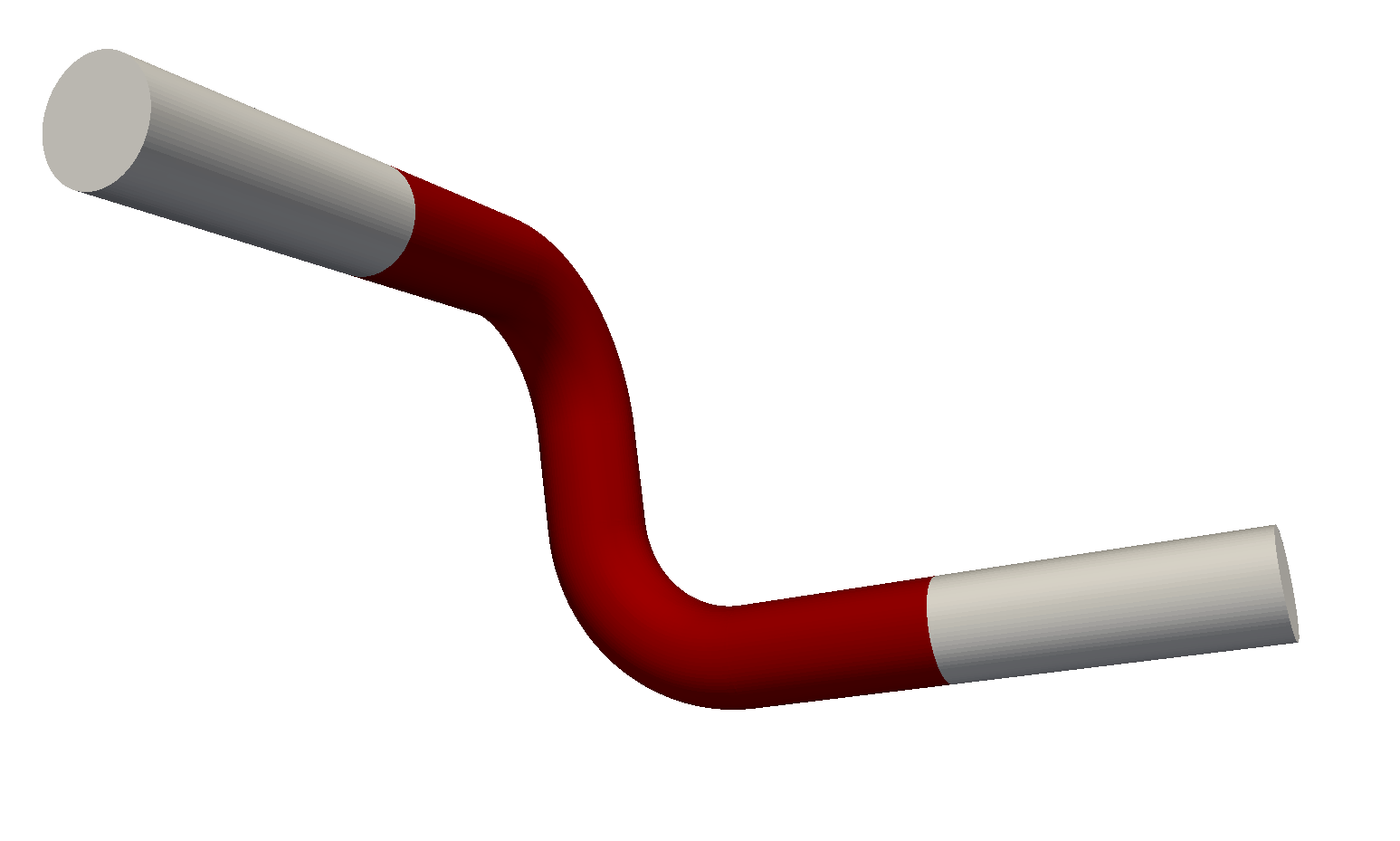}
}
\hspace{1cm}
\subfigure[]{
\centering
\includegraphics[scale=0.15]{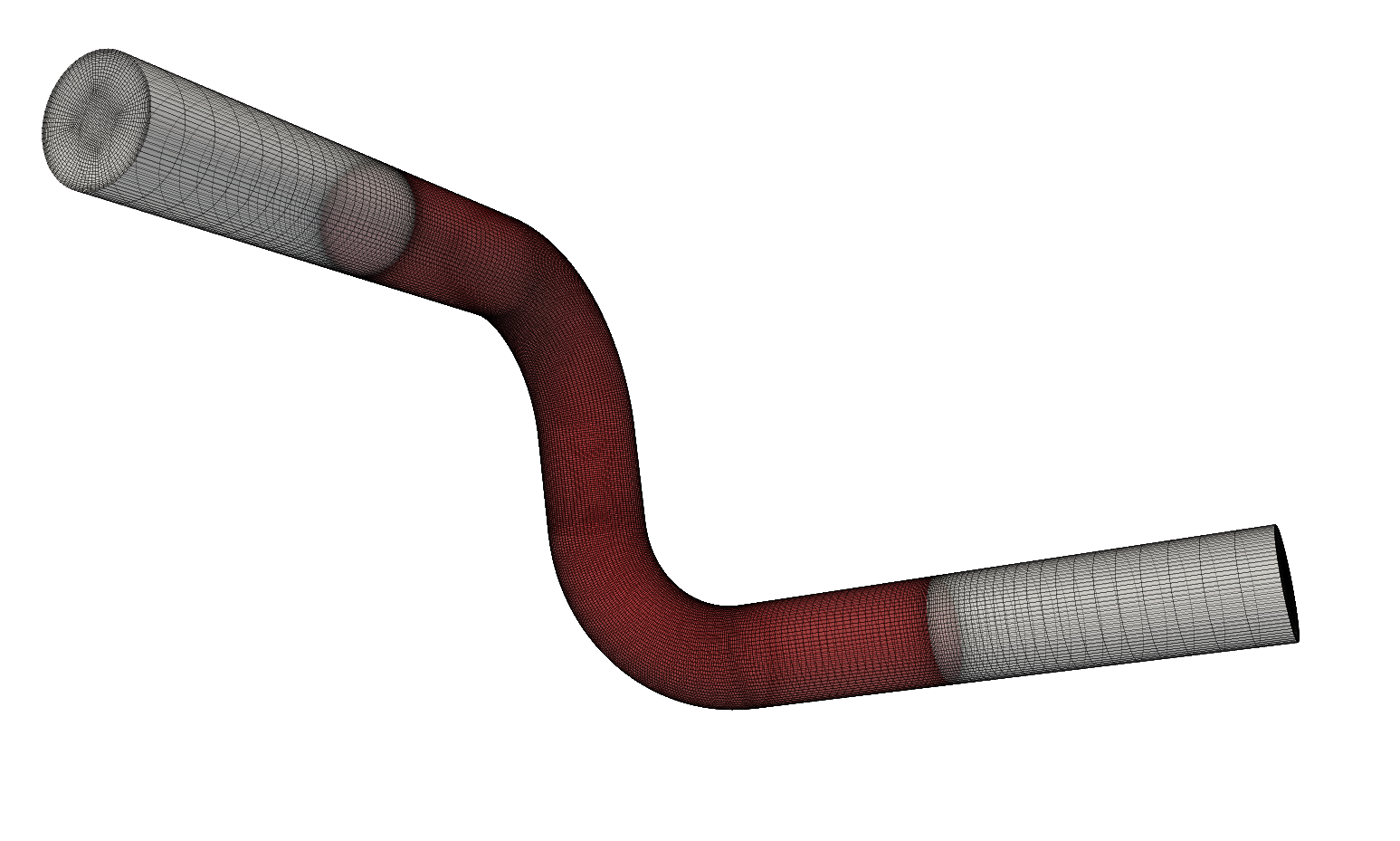}
}
\caption{Initial pipe geometry (a) as well as numerical grid (b)
 employed for the turbulent double bent pipe flow ($\mathrm{Re}_\mathrm{D} = 10^6$). Red areas indicate the design region.}
\label{fig:double_bent_pipe_perspective_initial}
\end{figure}

The final test case examines a three-dimensional double-bent pipe at a bulk Reynolds-number of $\mathrm{Re}_\mathrm{D} = U \, D / \nu = 10^5$,  where $U$, $D$ and $\nu$ refer to the bulk velocity, inlet diameter as well as the kinematic viscosity respectively, cf. Fig. \ref{fig:double_bent_pipe_1}.
A structured numerical grid of 820\,000 control volumes was used to mesh the internal flow field. Three diameters downstream of the inlet, the curved area is free for design in a CAD-free optimisation environment and discretized with 16\,000 surface elements as depicted in Fig. \ref{fig:double_bent_pipe_perspective_initial}. The grid is refined towards the transition between fixed and designed wall.
Along the inlet, a homogeneous velocity is imposed together with turbulent quantities that follow from the empirical relations (\ref{equ:turbulence_init}). A zero pressure value is prescribed at the outlet.
The ducted geometry is optimized w.r.t. the total power loss $J^\mathrm{P}$ outlined in  Eqn. (\ref{eqn:objectives2}). Hence, the adjoint boundary conditions coincide with those from the two-dimensional study in Sec. \ref{sec:tjunction}. 

\begin{figure}
\centering
\subfigure[]{
\centering
\includegraphics[scale=0.15]{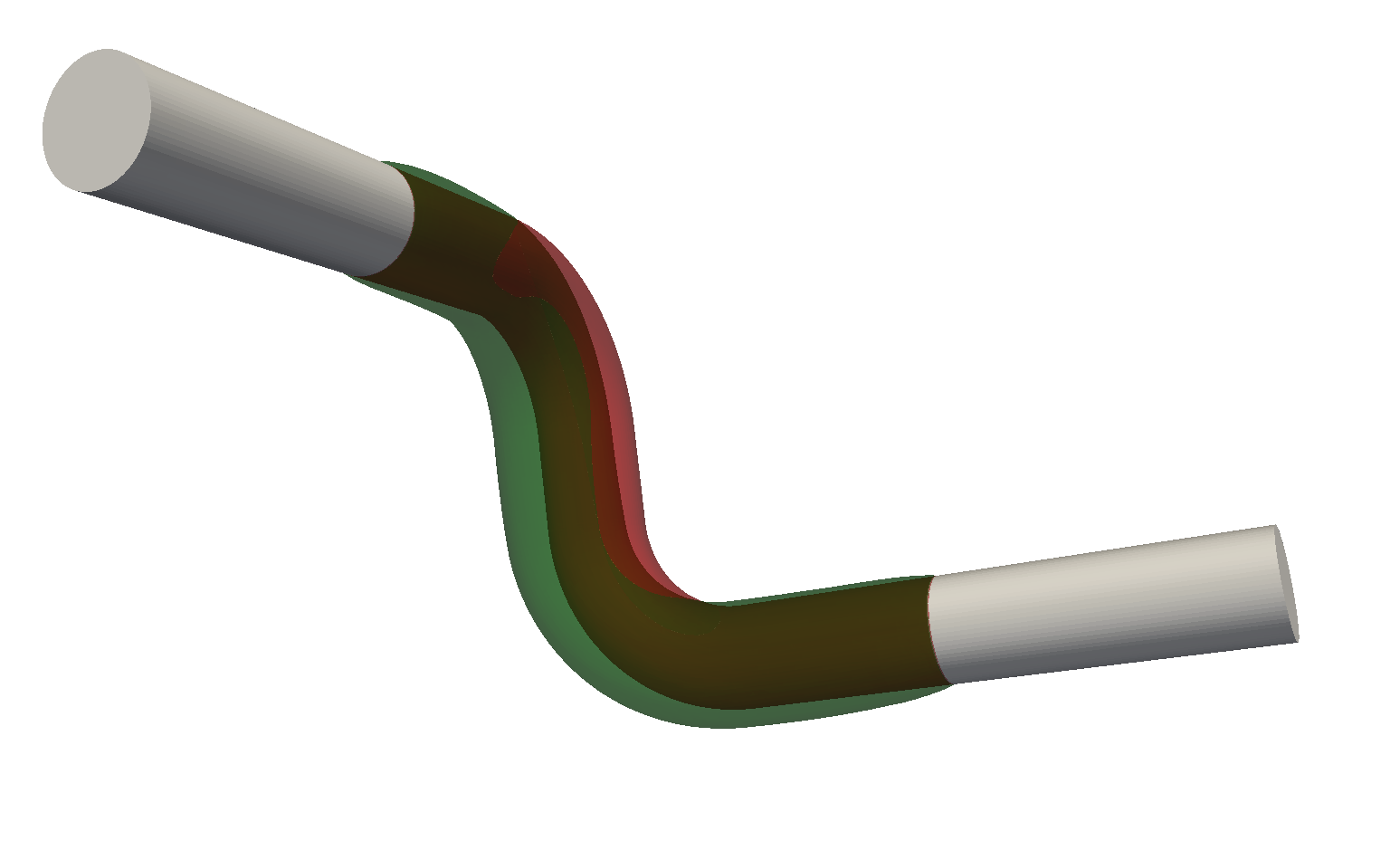}
}
\hspace{1cm}
\subfigure[]{
\centering
\analytiSolutionPictures
\begin{tikzpicture}
\begin{axis}[
 ylabel style={text width=0.25\textwidth,align=center},
 legend style={at={(0.98,0.98)},anchor=north east},
 xlabel={$n_{\mathrm{opt}}$ [-]},
 ylabel={$(J-J^{\mathrm{ini}})/J^{\mathrm{ini}} \cdot 100$ [\%]},
 xmin=0,xmax=40,
 ymin=-30,ymax=0,
]

\addplot [line2,mark1,each nth point=1] table [x expr={\thisrowno{0}},y expr={\thisrowno{1}}] {data/double_bent_pipe_frozen_opt.dat};
\addplot [line4,mark3,each nth point=1] table [x expr={\thisrowno{0}},y expr={\thisrowno{1}}] {data/double_bent_pipe_consistent_opt.dat};

\addlegendentry{F};
\addlegendentry{C};

\end{axis}
\end{tikzpicture}
}
\caption{Perspective view on the consistently optimized turbulent double bent pipe ($\mathrm{Re}_\mathrm{D} = 10^6$) (a) as well as 
the evolution of the power loss objective for the frozen turbulence (F) and the LoW-consistent (C) optimization framework.  
Red [green] areas indicate the initial [optimized] shape.}
\label{fig:double_bent_pipe_perspective_optimized}
\end{figure}

As illustrated by Fig. \ref{fig:double_bent_pipe_perspective_optimized}, the LoW-consistent framework (C) provides better convergence to 
an improved optimum when compared to the frozen turbulence approach (F). The respective differences are both significant and approximately amount to 22 \% improvements. 

\section{Conclusions}
\label{sec:conclusion}
The paper discussed the adjoint complement to the universal Law of the Wall (LoW) for fluid dynamic momentum boundary layers. The latter typically follows from a strongly simplified, unidirectional shear flow.
We first derived the adjoint companion of the simplified shear flow while distinguishing between two frequently used adjoint formulations. It is seen that both, the frozen turbulence strategy as well as a (differentiated) approach consistent to a mixing length model provide nearly the same adjoint equations. 
Moreover, the adjoint Law of the Wall essentially resembles the primal LoW, and the differences refer to a simple scaling with the ratio between the primal and the adjoint friction velocity
and the inclination in the logarithmic region, which reduces for the LoW-consistent approach.
The analysis displays that the  LoW-consistent approach is compatible with prominent RANS-type two-equation turbulence models, which ground on the mixing-length hypothesis.
Hence, an algebraic adjoint momentum closure can be formulated for the LoW which hooks up to any primal Boussinesq viscosity model
due to the assumed universal scaling of primal mean flow and turbulence quantities with the friction velocity.
%
The latter motivates a surprisingly simple algebraic turbulence treatment for the adjoint momentum equation.
This LoW-consistent formulation is expressed by halving the velocity inclination  entering a wall function boundary condition in the logarithmic region and doubling the turbulent viscosity. Comparing to the frozen turbulence approach, the LoW-consistent method is deemed a better approximation for adjoint flow optimisation efforts.
%
%
Results obtained by the LoW-consistent algebraic closure come at no extra cost and indicate an acceleration of the optimization process as well as improved optimal solutions for 
shape optimizations of external and internal engineering flows.
A hidden benefit of the suggested LoW-consistent approach refers to the enhanced stability of the numerical framework due to the augmented viscosity.

\section{Acknowledgments}
The current work is a part of the research projects "Drag Optimisation of Ship Shapes’" funded by the German Research Foundation (DFG, Grant No. RU 1575/3-1)
as well as "Dynamic Adaptation of Modular Shape Optimization Processes" funded by the German Federal Ministry for Economic Affairs and Energy (BMWi, Grant No.  03SX453B).
The research takes places within the Research Training Group (RTG) 2583 "Modeling, Simulation and Optimization with Fluid Dynamic Applications" funded by the German Research Foundation.
This support is gratefully acknowledged by the authors.
Selected computations were performed with resources provided by the North-German Super-computing Alliance (HLRN).

\section{Authorship Contribution Statement}
\textbf{Niklas K{\"u}hl}: Conceptualization, Methodology, Software, Validation, Formal analysis, Investigation, Writing - original draft, Visualization, Writing - review \& editing.
\textbf{Peter M. M\"uller}: Formal analysis, Writing - review \& editing.
\textbf{Thomas Rung}: Project administration, Funding acquisition, Supervision, Formal analysis, Investigation, Writing - original draft, Writing - review \& editing.

\section{Further Declarations}

\subsection{Conflicts of Interest}
The authors declare that they have no conflict of interest.

\subsection{Availability of Data and Material}
Not applicable

\subsection{Code Availability}
Not applicable

\bibliographystyle{plain}
\bibliography{library}

\begin{thebibliography}{10}

\bibitem{bagheri2020adjoint}
A.K. Bagheri and A.~Da~Ronch.
\newblock Adjoint-{B}ased {S}urrogate {M}odelling of {S}palart-{A}llmaras
  {T}urbulence {M}odel {U}sing {G}radient {E}nhanced {K}riging.
\newblock In {\em AIAA AVIATION 2020 FORUM}, 2020.

\bibitem{bueno2012continuous}
A.~Bueno-Orovio, C.~Castro, F.~Palacios, and E.~Zuazua.
\newblock Continuous {A}djoint {A}pproach for the {S}palart-{A}llmaras {M}odel
  in {A}erodynamic {O}ptimization.
\newblock {\em AIAA Journal}, 50(3):631--646, 2012.

\bibitem{dwight2006effects}
R.~Dwight and J.~Br{\'e}zillon.
\newblock Effects of {V}arious {A}pproximations of the {D}iscrete {A}djoint on
  {G}radient-{B}ased {O}ptimization.
\newblock {\em AIAA paper 2006}, 690, 2006.

\bibitem{giles1997adjoint}
M.B. Giles and N.A. Pierce.
\newblock Adjoint {E}quations in {CFD}: {D}uality, {B}oundary {C}onditions and
  {S}olution {B}ehaviour.
\newblock {\em AIAA Paper}, 1997.
\newblock AIAA--97--1850.

\bibitem{giles2000introduction}
M.B. Giles and N.A. Pierce.
\newblock An {I}ntroduction to the {A}djoint {A}pproach to {D}esign.
\newblock {\em Flow, Turbulence and Combustion}, 65(3):393--415, 2000.

\bibitem{hartmann2011adjoint}
R.~Hartmann, J.~Held, and T.~Leicht.
\newblock Adjoint-{B}ased {E}rror {E}stimation and {A}daptive {M}esh
  {R}efinement for the {RANS} and k-$\omega$ {T}urbulence {M}odel {E}quations.
\newblock {\em Journal of Computational Physics}, 230(11):4268--4284, 2011.

\bibitem{hucho2002aerodynamik}
W.H. Hucho.
\newblock {\em Aerodynamik der {S}tumpfen {K}{\"o}rper}.
\newblock Springer, 2002.

\bibitem{jones1972prediction}
W.P. Jones and B.E. Launder.
\newblock The {P}rediction of {L}aminarization with a {T}wo-{E}quation {M}odel
  of {T}urbulence.
\newblock {\em International Journal of Heat and Mass Transfer},
  15(2):301--314, 1972.

\bibitem{kapellos2019unsteady}
C.S. Kapellos, E.M. Papoutsis-Kiachagias, K.C. Giannakoglou, and M.~Hartmann.
\newblock The {U}nsteady {C}ontinuous {A}djoint {M}ethod for {M}inimizing
  {F}low-{I}nduced {S}ound {R}adiation.
\newblock {\em Journal of Computational Physics}, 392:368--384, 2019.

\bibitem{kavvadias2015continuous}
I.S. Kavvadias, E.M. Papoutsis-Kiachagias, G.~Dimitrakopoulos, and K.C.
  Giannakoglou.
\newblock The {C}ontinuous {A}djoint {A}pproach to the k--$\omega$ {SST}
  {T}urbulence {M}odel with {A}pplications in {S}hape {O}ptimization.
\newblock {\em Engineering Optimization}, 47(11):1523--1542, 2015.

\bibitem{kroger2018adjoint}
J.~Kr{\"o}ger, N.~K{\"u}hl, and T.~Rung.
\newblock Adjoint {V}olume-of-{F}luid {A}pproaches for the {H}ydrodynamic
  {O}ptimisation of {S}hips.
\newblock {\em Ship Technology Research}, 65(1):47--68, January 2018.

\bibitem{kroger2015cad}
J.~Kr{\"o}ger and T.~Rung.
\newblock {CAD}-{F}ree {H}ydrodynamic {O}ptimisation {U}sing {C}onsistent
  {K}ernel-{B}ased {S}ensitivity {F}iltering.
\newblock {\em Ship Technology Research}, 62(3):111--130, 2015.

\bibitem{kuhl2020adjoint}
N.~K{\"u}hl, J.~Kr{\"o}ger, M.~Siebenborn, M.~Hinze, and T.~Rung.
\newblock Adjoint {C}omplement to the {V}olume-of-{F}luid {M}ethod for
  {I}mmiscible {F}lows.
\newblock {\em arXiv preprint arXiv:2009.03957}, 2020.

\bibitem{kuhl2020continuous}
N.~K{\"u}hl, P.~M. M{\"u}ller, and T.~Rung.
\newblock Continuous {A}djoint {C}omplement to the {B}lasius {E}quation.
\newblock {\em arXiv preprint arXiv:2011.07583}, 2020.

\bibitem{kuhl2019decoupling}
N.~K{\"u}hl, P.~M. M{\"u}ller, A.~St{\"u}ck, M.~Hinze, and T.~Rung.
\newblock Decoupling of {C}ontrol and {F}orce {O}bjective in {A}djoint-{B}ased
  {F}luid {D}ynamic {S}hape {O}ptimization.
\newblock {\em AIAA Journal}, 57(9):4110--4114, 2019.

\bibitem{rung2001universal}
H.M. L{\"u}bcke, T.~Rung, and F.~Thiele.
\newblock Universal {W}all-{B}oundary {C}onditions for {T}urbulence {T}ransport
  {M}odels.
\newblock {\em Journal of Applied Mathematics and Mechanics}, 81:481--482,
  2001.

\bibitem{manservisi2016numerical}
S.~Manservisi and F.~Menghini.
\newblock Numerical {S}imulations of {O}ptimal {C}ontrol {P}roblems for the
  {R}eynolds {A}veraged {N}avier-{S}tokes {S}ystem {C}losed with a
  {T}wo-{E}quation {T}urbulence {M}odel.
\newblock {\em Computers \& Fluids}, 125:130--143, 2016.

\bibitem{manservisi2016optimal}
S.~Manservisi and F~Menghini.
\newblock Optimal {C}ontrol {P}roblems for the {N}avier--{S}tokes {S}ystem
  {C}oupled with the k-$\omega$ {T}urbulence {M}odel.
\newblock {\em Computers \& Mathematics with Applications}, 71(11):2389--2406,
  2016.

\bibitem{marta2013handling}
A.C. Marta and S.~Shankaran.
\newblock On the {H}andling of {T}urbulence {E}quations in {RANS} {A}djoint
  {S}olvers.
\newblock {\em Computers \& Fluids}, 74:102--113, 2013.

\bibitem{nielsen2013discrete}
E.J. Nielsen and B.~Diskin.
\newblock Discrete {A}djoint-{B}ased {D}esign for {U}nsteady {T}urbulent
  {F}lows on {D}ynamic {O}verset {U}nstructured {G}rids.
\newblock {\em AIAA journal}, 51(6):1355--1373, 2013.

\bibitem{nielsen2010discrete}
E.J. Nielsen, B.~Diskin, and N.K. Yamaleev.
\newblock Discrete {A}djoint-{B}ased {D}esign {O}ptimization of {U}nsteady
  {T}urbulent {F}lows on {D}ynamic {U}nstructured {G}rids.
\newblock {\em AIAA journal}, 48(6):1195--1206, 2010.

\bibitem{nielsen2004implicit}
E.J. Nielsen, J.~Lu, M.A. Park, and D.L. Darmofal.
\newblock An {I}mplicit, {E}xact {D}ual {A}djoint {S}olution {M}ethod for
  {T}urbulent {F}lows on {U}nstructured {G}rids.
\newblock {\em Computers \& Fluids}, 33(9):1131--1155, 2004.

\bibitem{othmer2008continuous}
C.~Othmer.
\newblock A {C}ontinuous {A}djoint {F}ormulation for the {C}omputation of
  {T}opological and {S}urface {S}ensitivities of {D}ucted {F}lows.
\newblock {\em International Journal for Numerical Methods in Fluids},
  58(8):861--877, 2008.

\bibitem{othmer2014adjoint}
C.~Othmer.
\newblock Adjoint {M}ethods for {C}ar {A}erodynamics.
\newblock {\em Journal of Mathematics in Industry}, 4(1):6, 2014.

\bibitem{papoutsis2016continuous}
E.M. Papoutsis-Kiachagias and K.C. Giannakoglou.
\newblock Continuous {A}djoint {M}ethods for {T}urbulent {F}lows, {A}pplied to
  {S}hape and {T}opology {O}ptimization: {I}ndustrial {A}pplications.
\newblock {\em Archives of Computational Methods in Engineering}, 23(2):255,
  2016.

\bibitem{papoutsis2015continuous}
E.M. Papoutsis-Kiachagias, A.S. Zymaris, I.S. Kavvadias, D.I. Papadimitriou,
  and K.C. Giannakoglou.
\newblock The {C}ontinuous {A}djoint {A}pproach to the k-{$\epsilon$}
  {T}urbulence {M}odel for {S}hape {O}ptimization and {O}ptimal {A}ctive
  {C}ontrol of {T}urbulent {F}lows.
\newblock {\em Engineering Optimization}, 47(3):370--389, 2015.

\bibitem{pope2001turbulent}
S.B. Pope.
\newblock Turbulent {F}lows, 2001.

\bibitem{prandtl25}
L.~Prandtl.
\newblock Bericht \"uber die {E}ntstehung der {T}urbulenz.
\newblock {\em Zeitschrift f{\"u}r Angewandte Mathematik und Mechanik},
  5:136--139, 1925.

\bibitem{rung2009challenges}
T.~Rung, K.~W{\"o}ckner, M.~Manzke, J.~Brunswig, C.~Ulrich, and A.~St{\"u}ck.
\newblock Challenges and {P}erspectives for {M}aritime {CFD} {A}pplications.
\newblock {\em Jahrbuch der Schiffbautechnischen Gesellschaft}, 103:127--39,
  2009.

\bibitem{schulz2016computational}
V.~Schulz and M.~Siebenborn.
\newblock Computational {C}omparison of {S}urface {M}etrics for {PDE}
  {C}onstrained {S}hape {O}ptimization.
\newblock {\em Computational Methods in Applied Mathematics}, 16(3):485--496,
  2016.

\bibitem{soto2004computation}
O.~Soto and R.~L{\"o}hner.
\newblock On the {C}omputation of {F}low {S}ensitivities from {B}oundary
  {I}ntegrals.
\newblock In {\em 42 nd AIAA Aerospace Sciences Meeting and Exhibit}, 2004.

\bibitem{soto2004adjoint}
O.~Soto, R.~L{\"o}hner, and C.~Yang.
\newblock An {A}djoint-{B}ased {D}esign {M}ethodology for {CFD} {P}roblems.
\newblock {\em International Journal of Numerical Methods for Heat \& Fluid
  Flow}, 14(6):734--759, 2004.

\bibitem{spalart1992one}
P.~Spalart and S.~Allmaras.
\newblock A {O}ne-{E}quation {T}urbulence {M}odel for {A}erodynamic {F}lows.
\newblock In {\em 30th AIAA Aerospace Sciences Meeting and Exhibit}, page 439,
  1992.

\bibitem{stuck2011adjoint}
A.~St{\"u}ck and T.~Rung.
\newblock Adjoint {RANS} with {F}iltered {S}hape {D}erivatives for
  {H}ydrodynamic {O}ptimisation.
\newblock {\em Computers \& Fluids}, 47(1):22--32, 2011.

\bibitem{stuck2013adjoint}
A.~St{\"u}ck and T.~Rung.
\newblock Adjoint {C}omplement to {V}iscous {F}inite-{V}olume
  {P}ressure-{C}orrection {M}ethods.
\newblock {\em Journal of Computational Physics}, 248:402--419, 2013.

\bibitem{vanDriest1956turbulent}
E.R. Van~Driest.
\newblock On {T}urbulent {F}low {N}ear a {W}all.
\newblock {\em Journal of the Aeronautical Sciences}, 23(11):1007--1011, 1956.

\bibitem{vassberg2006aerodynamic_II}
J.~Vassberg and A.~Jameson.
\newblock Aerodynamic {S}hape {O}ptimization {P}art {2}: {S}ample
  {A}pplications.
\newblock {\em Introduction to Optimization and Multidisciplinary Design},
  pages 1--41, 2006.

\bibitem{vassberg2006aerodynamic}
J.~Vassberg and A.~Jameson.
\newblock Aerodynamic {S}hape {O}ptimization {P}art {I}: {T}heoretical
  {B}ackground.
\newblock {\em Introduction to Optimization and Multidisciplinary Design},
  pages 1--30, 2006.

\bibitem{wilcox1998turbulence}
D.C. Wilcox.
\newblock {\em Turbulence {M}odeling for {CFD}}, volume~2.
\newblock DCW Industries La Canada, 1998.

\bibitem{yakubov2013hybrid}
S.~Yakubov, B.~Cankurt, M.~Abdel-Maksoud, and T.~Rung.
\newblock Hybrid {MPI}/{O}pen{MP} {P}arallelization of an {E}uler-{L}agrange
  {A}pproach to {C}avitation {M}odelling.
\newblock {\em Computers \& Fluids}, 80:365--371, 2013.

\bibitem{yakubov2015experience}
S.~Yakubov, T.~Maquil, and T.~Rung.
\newblock Experience {U}sing {P}ressure-{B}ased {CFD} {M}ethods for
  {Euler}-{Euler} {S}imulations of {C}avitating {F}lows.
\newblock {\em Computers \& Fluids}, 111:91--104, 2015.

\bibitem{zymaris2009continuous}
A.S. Zymaris, D.I. Papadimitriou, K.C. Giannakoglou, and C.~Othmer.
\newblock Continuous {A}djoint {A}pproach to the {S}palart-{A}llmaras
  {T}urbulence {M}odel for {I}ncompressible {F}lows.
\newblock {\em Computers \& Fluids}, 38(8):1528--1538, 2009.

\bibitem{zymaris2010adjoint}
A.S. Zymaris, D.I. Papadimitriou, K.C. Giannakoglou, and C.~Othmer.
\newblock Adjoint {W}all {F}unctions: {A} {N}ew {C}oncept for {U}se in
  {A}erodynamic {S}hape {O}ptimization.
\newblock {\em Journal of Computational Physics}, 229(13):5228--5245, 2010.

\end{thebibliography}

\end{document}